\documentclass[aps,prb,twocolumn,showpacs,amsmath,amssymb,superscriptaddress]{revtex4-2}
\usepackage{graphicx}
\usepackage{CJK}
\usepackage{color}
\usepackage[colorlinks,bookmarks=false,citecolor=blue,linkcolor=red,urlcolor=blue]{hyperref}
\usepackage{multirow}
\usepackage{ulem}
\usepackage{tabularx}
\usepackage[nounderscore]{syntax}
\graphicspath{{figs/}}
\usepackage{subfigure}
\usepackage{float}

\newcommand{\be}{\begin{equation}}
\newcommand{\ee}{\end{equation}}

\begin{document}
\begin{CJK*}{UTF8}{gbsn}
\title{Delocalization of skin steady states}

\author{Xu Feng}
\affiliation{Beijing National Laboratory for Condensed Matter Physics, Institute
of Physics, Chinese Academy of Sciences, Beijing 100190, China}
\affiliation{School of Physical Sciences, University of Chinese Academy of Sciences,
Beijing 100049, China }

\author{Shu Chen}
\email{schen@iphy.ac.cn }
\affiliation{Beijing National Laboratory for Condensed Matter Physics, Institute
of Physics, Chinese Academy of Sciences, Beijing 100190, China}
\affiliation{School of Physical Sciences, University of Chinese Academy of Sciences,
Beijing 100049, China }
% \affiliation{Yangtze River Delta Physics Research Center, Liyang, Jiangsu 213300,
% China }

\date{\today}
\begin{abstract}
The skin effect, characterized by the tendency of particles to accumulate at the boundaries, has been extensively studied in non-Hermitian systems. In this work, we propose an intuitive Lindbladian composed of two chains with reversed skin localization. The skin steady state is gradually delocalized as the interchain coupling increases.  
In the single-body scenario, it corresponds to a shift in the scaling of the Liouvillian gap $\Delta$ from $\Delta \propto N^0$ to $\Delta \propto N^{-2}$. Notably, exact diagonalization results reveal a system-size sensitivity of the single-particle Liouvillian spectrum, inherited from the non-Hermitian effective Hamiltonian's system-size sensitivity. We predict that even an arbitrarily small coupling will induce dramatic changes in the Liouvillian spectrum and steady state in the thermodynamic limit, a phenomenon we term the critical Liouvillian skin effect. Additionally, in the many-body scenario, by employing the stochastic Schr\"odinger equation to unravel the Lindblad master equation, it is revealed that the scaling behavior of steady-state entanglement changes from the area law to the logarithmic law.
This work demonstrates the delocalization of both single-body and many-body skin steady states, introducing a novel mechanism for inducing entanglement transitions beyond the quantum Zeno effect.

\end{abstract}
\maketitle

\section{Introduction}
\label{intro}

In past years, numerous unique and novel phenomena have been discovered in non-Hermitian systems \cite{ashida2020non}, one of which is the non-Hermitian skin effect (NHSE) \cite{GBZ}. The non-Hermitian systems exhibiting NHSE are extremely sensitive to boundary conditions \cite{GBZ,boundarysensitive}. When switching from periodic boundary conditions to open boundary conditions, a substantial number of Bloch waves become localized at the boundaries, accompanied by significant changes in the energy spectrum. In addition, the topological origin of NHSE has been elucidated \cite{WindingSkin,windingSkin2,Slager}, and the generalized Brillouin zone (GBZ) theory is developed to analytically characterize one-dimensional NHSE in the thermodynamic limit \cite{GBZ,non-bloch,auxiliaryGBZ}. Recently, NHSE has also been extended into higher-dimension, revealing more complex and richer properties \cite{zhangUniversalNonHermitianSkin2022,DimensionalTransmutation,Amoeba,huNonHermitianBandTheory2023,huNHSEhighDimensions}.
However, according to the quantum trajectory theory \cite{MoteCarloWavefunction,quantumtrajectory}, on the one hand, the non-Hermitian Hamiltonian (NHH) requires post-selection, rendering it exponentially difficult to realize experimentally; on the other hand, quantum jumps will inevitably modify or even invalidate the physical pictures provided by the NHH \cite{LSEFermionBoson,NHfluctuation}. Therefore, it is more practical to investigate the skin effect using the Lindbladian formalism rather than NHH. Indeed, the skin effect has already been generalized into the open quantum systems (OQS) recently,  dubbed as the Liouvillian skin effect (LSE). It is found that LSE likewise displays boundary-sensitivity \cite{LSEFermionBoson,BoundarySensitiveLindb,ManyBodyNHSEgaugecoupling2023,hagaLiouvillianSkinEffect2021,mao-liouvillian2024} and plays a nontrivial role in the relaxation dynamics \cite{Liouvillianskin,chiraldamping,helicaldamping,hagaLiouvillianSkinEffect2021,BoundarySensitiveLindb,wangAcceleratingRelaxationDynamics2023,symmetrizedLiouvillianGap,Phantomrelaxation}, quantum criticality \cite{QuantumCriticalitynonreciprocity}, quantum transport \cite{BosonicLSE}, etc. 

Specifically, it is shown that proper continuous measurement with conditional feedback can achieve the LSE in the steady state, thereby freeing the particles and driving the system into the area law phase, irrespective of measurement strength \cite{MISE,MISE-longrange}. So it is interesting to design a setup in which the LSE is adjustable and explore whether the scaling behavior of entanglement will vary when the LSE is progressively diminished. 
Moreover, to the best of our knowledge, 
the delocalization of skin steady states for the full Lindbladian has been scarcely studied and it is restricted only in the single-particle scenario \cite{NHfluctuation}. 
Therefore, to uncover more features of this delocalization process, we propose an intuitive setup involving two chains with reversed Liouvillian skin localization, where the coupling strength between these chains is gradually increased. 
Correspondingly, both for the single-body and many-body cases, the skin localization of the non-equilibrium steady state diminishes progressively and eventually disappears,  
witnessed by various dynamic signatures such as the shift in the scaling of the Liouvillian gap and entanglement entropy.
It is worth mentioning that delocalization of skin steady state acts as a novel mechanism for entanglement transition, distinct from the widely studied measurement-induced phase transition (MIPT) \cite{Nahum,FisherPRB1,FishePRB2,fermionMIPT,DiehlMIPT,longrangeMIPT,MIPTtwodimension,MIPTLiu1,MIPTLiu2,MIPTLiu3} and Anderson localization transition \cite{EPTAnderson,EPTLocalizationlongRange}. Additionally, our model is also capable of attaining MIPT in the extended regime, as discussed later.
At last, it is crucial to emphasize that our work differs from previous studies of entanglement transition caused by NHSE, which corresponds to the limit without quantum jumps and relies on postselection \cite{EPTNonHermitian,liDisorderInducedEntanglementPhase2023,NHEvoQuasiDisorder,NHEvoAA,liu-nonlinear-2024}. 

\section{Model}

\begin{figure}[htb]
\includegraphics[width=0.48\textwidth]{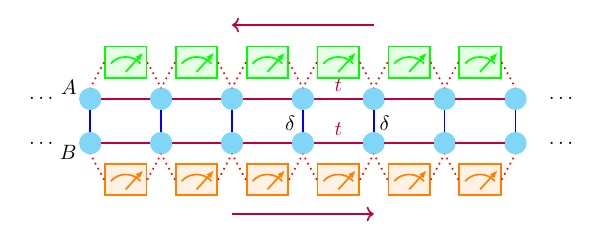}
\caption{The Schematic diagram for the setup. For chain $A$ ($B$), the Lindblad operators (continuous measurement with conditional feedback) tend to drive the particles to the left (right) side. The intrachain hopping amplitude is $t$ and the interchain hopping amplitude is $\delta$. The open boundary conditions are adopted. }
\label{fig1}
\end{figure}

Under Markovian approximation, the evolution of the system's density matrix $\rho$  obeys the Lindblad master equation (LME):
\begin{equation}\label{LME}
\begin{aligned}
\dfrac{d\rho(t)}{dt}=\mathcal{L}[\rho(t)]=-\mathrm{i}[H_0, \rho(t)]+\sum_{\nu=\{A,B\}}\sum_{j=1}^{N-1}\mathcal{D}_{j,\nu}[\rho(t)],
\end{aligned}
\end{equation}
where the dissipators $\mathcal{D}_{j,\nu}[\rho]$ are defined as
$\mathcal{D}_{j,\nu}[\rho]=L_{j,\nu}\rho L_{j,\nu}^{\dagger}-\dfrac{1}{2}\{L_{j,\nu}^{\dagger}L_{j,\nu},\rho\}$. The Lindblad operators acting on chain $A$ and chain $B$ are given by $L_{j,A}=e^{\mathrm{i}\pi n_{j+1,A}}\xi_{j,A}^{\dagger}\xi_{j,A}$
and $L_{j,B}=e^{\mathrm{i}\pi n_{j+1,B}}\xi_{j,B}^{\dagger}\xi_{j,B}$, respectively, in which $\xi_{j,A}=(c_{j,A}+\mathrm{i}c_{j+1,A})/\sqrt{2}$ and $\xi_{j,B}=(c_{j,B}-\mathrm{i}c_{j+1,B})/\sqrt{2}$. Physically, it is known that $L_{j,\nu}$ can be understood as the combination of continuous measurements $\xi_{j,\nu}^{\dagger}\xi_{j,\nu}$ and subsequent conditional feedback $e^{\mathrm{i}\pi n_{j+1,\nu}}$, and $L_{j,A}$ ($L_{j,B}$) tends to drive the particles into the left (right) side \cite{MISE,MISE-longrange,MISE-DPT,BoundarySensitiveLindb}.
For the sake of simplicity, we consider the dissipation (or measurement) strength to be uniform and denoted as $\gamma$. Unless otherwise specified,
we choose $\gamma=0.5$ hereafter.

The system Hamiltonian $H_0$ describes a non-interacting two-leg ladder model:
\begin{equation}
\begin{aligned}
H_{0}&=t\sum_{j=1}^{N-1}(c_{j,A}^{\dagger}c_{j+1,A}+c_{j,B}^{\dagger}c_{j+1,B}+\text{h.c.}) \\
&+\delta\sum_{j=1}^N (c^{\dagger}_{j,A}c_{j,B}+\text{h.c.}),
\end{aligned}
\end{equation}
in which $t$ is intrachain hopping strength (we set $t=1$ throughout the work), and $\delta$ denotes the tunable interchain hopping strength. The whole setup is displayed schematically in Fig. \ref{fig1}.
Henceforth,  the effective non-Hermitian Hamiltonian is 
$H_{\text{eff}}=H_{0}-\dfrac{\mathrm{i}}{2}\gamma\sum_{\nu=\{A,B\}}\sum_{j=1}^{N-1}L_{j,\nu}^{\dagger}L_{j,\nu} =\sum_{j=1}^{N-1}(t+\dfrac{\gamma}{4})c_{j,A}^{\dagger}c_{j+1,A}+(t-\dfrac{\gamma}{4})c_{j+1,A}^{\dagger}c_{j,A}
+(t-\dfrac{\gamma}{4})c_{j,B}^{\dagger}c_{j+1,B} +
(t+\dfrac{\gamma}{4})c_{j+1,B}^{\dagger}c_{j,B}
 -\dfrac{\mathrm{i}}{4}\gamma(n_{j,A}+n_{j+1,A}+n_{j,B}+n_{j+1,B}) + \delta\sum_{j=1}^N (c^{\dagger}_{j,A}c_{j,B}+ c^{\dagger}_{j,B}c_{j,A})$, which represents a coupled Hatano-Nelson model with onsite imaginary potential  \cite{HNmodel}. The similar non-Hermitian effective Hamiltonian $H_{\text{eff}}$ is found to exhibit critical non-Hermitian skin effect (CNHSE) \cite{CNHSE,scalingCNHSE}, which imprints itself in the Liouvillian spectrum as shown later.

\section{single-body case}
\label{single-body case}

\begin{figure}[htb]
\includegraphics[width=0.49\textwidth]{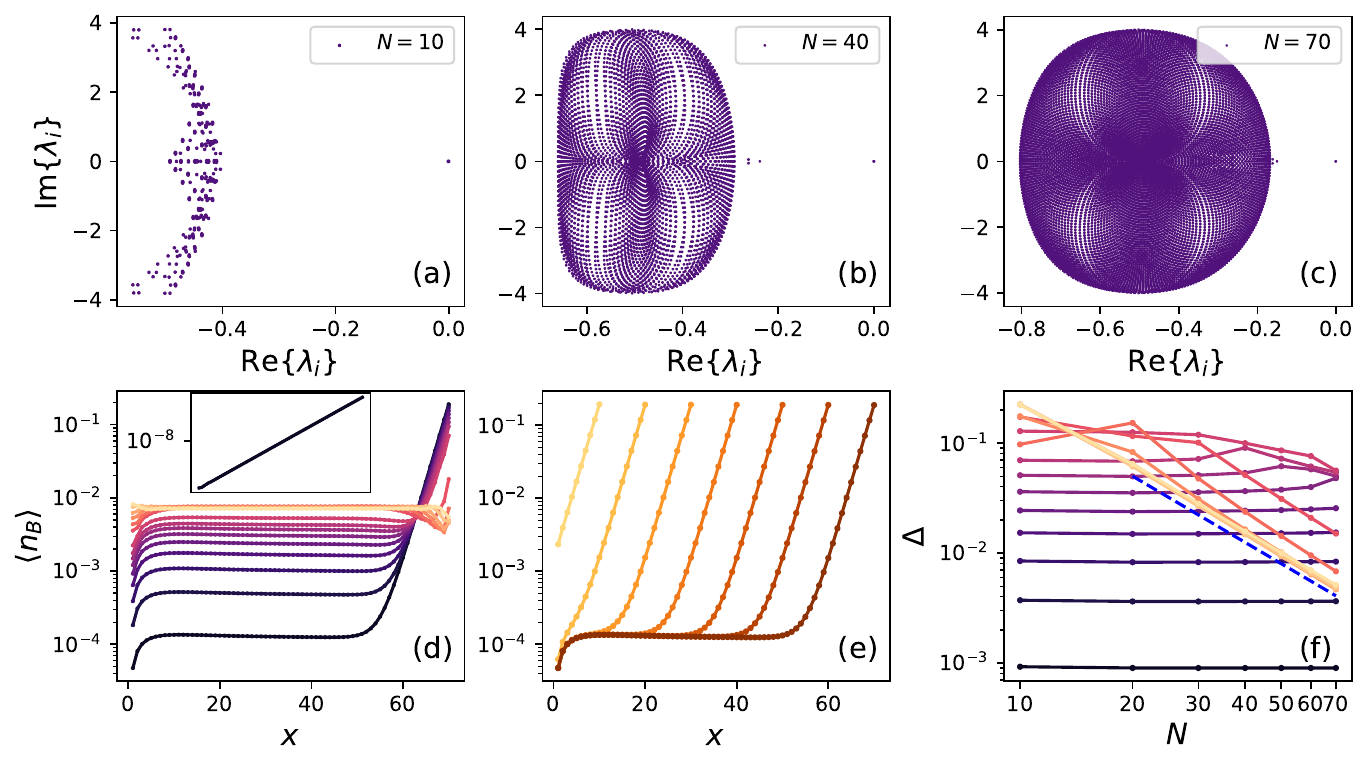}
\caption{$\gamma=0.5$, (a),(b),(c) Liouvillian spectra. $\delta=0.01$, $N=10,40,70$, respectively. (d) The steady-state density distribution of chain $B$ for various interchain hopping $\delta$ in the semi-log plot. $N=70$. The inset corresponds to zero coupling $\delta=0.0$, the main plot corresponds to nonzero coupling.
(e) The steady-state density distribution of chain $B$ for various system sizes $N$ in the semi-log plot. $\delta=0.01$, from the lightest color to darkest color, $N=10, 20, 30, 40, 50, 60, 70$. 
(f) The scaling of Liouvillian gap $\Delta$ with respect to $N$ for various $\delta$ in the semi-log plot. The Liouvillian gap changes from $\Delta\propto N^0$ to $N^{-2}$ with increasing interchain hopping $\delta$. 
In (d) (main plot) and (f), from the darkest black color to the lightest yellow color, the interchain hopping is $\delta=0.01, 0.02, 0.03, 0.04, 0.05, 0.06, 0.07, 0.08, 0.1, 0.2, 0.3, 0.4,$ $0.5, 0.7, 1.0.$ }
\label{fig2}
\end{figure}

The total particle number of our system is conserved due to the strong $U(1)$ symmetry of the Lindbladian, as proved by the commutation relations $[{N}_{\text{tot}}, H_0]=[{N}_{\text{tot}}, L_{j,\nu}]=0$, where ${N}_{\text{tot}}=\sum_{\nu}\sum_{j=1}^Nc_{j,\nu}^{\dagger}c_{j,\nu}$.
Therefore, we first focus on the single-particle case for simplicity. Moreover, besides the strong $U(1)$ symmetry, the single-body Lindbladian also shares rotation symmetry. Defining the rotation operation $\mathcal{P}$ with respect to the center of the ladder such that $\mathcal{P}c_{j,A}\mathcal{P}^{-1}=c_{N+1-j,B}$, it can be verified that  $\mathcal{P} H_{0}\mathcal{P}^{-1}=H_0$ and $\mathcal{P}L_{j,A}\mathcal{P}^{-1}=-L_{N-j,B}$. Consequently, $\mathcal{P}\mathcal{L}\mathcal{P}^{-1}=\mathcal{L}$ holds, implying that the steady state adheres to $\langle n_{j,A}\rangle=\langle n_{N+1-j,B} \rangle$. However, in the many-body case, the interaction term $n_{j,\nu}n_{j+1,\nu}$ in the Lindblad operator $L_{j,\nu}$ breaks this symmetry, leading to $\langle n_{j,A}\rangle\neq \langle n_{N+1-j,B}\rangle$. In the following, 
we numerically obtain the Liouvillian spectra, Liouvillian gap $\Delta$, and steady state by exact diagonalization.

First of all,  the Liouvillian spectra are sensitive to system size. As denoted in Fig. \ref{fig2}(a),(b),(c), the range of the real part of the Liouvillian spectra gradually expands with increasing system size $N$.  This size sensitivity originates from the system-size sensitivity of 
the non-Hermitian effective Hamiltonian $H_{\text{eff}}$. As detailed in the Appendix. \ref{sec:appendixA}, the zeroth-order perturbation results already reproduce the system-size sensitivity, which unveils the close connection between LSE and NHSE for our Lindbladian. Generally speaking, it provides a routine to construct a Lindbladian with LSE by designing a non-Hermitian effective Hamiltonian with NHSE. Moreover, according to the auxiliary generalized Brillouin zone theory, any nonzero coupling leads to the abrupt change of the generalized Brillouin zone, resulting in a dramatic change of energy spectrum of $H_{\text{eff}}$ in the thermodynamic limit, which is named critical non-Hermitian skin effect  \cite{CNHSE,auxiliaryGBZ,scalingCNHSE}. Consequently, based on the perturbation analysis, we can predict that arbitrary small interchain hopping $\delta$ will also cause a sharp change in the Liouvillian spectrum and steady state in the thermodynamic limit, which is analogously dubbed the critical Liouvillian skin effect [ see Appendix. \ref{sec:appendixA}].

Secondly, as shown in the inset of Fig. \ref{fig2}(d), if the chains are decoupled, i.e. $\delta=0$, the steady-state density in chain $B$ ($\langle n_B\rangle$) decays exponentially through the whole chain, similar to the localization of the Hatano-Nelson model. When slightly turning on the interchain hopping $\delta$, 
for example, $\delta=0.01$, $\langle n_B\rangle$ remains right skin-localized. However, the exponential localization behavior only persists within finite width $\zeta$ (denoted as the skin region) but not the whole chain anymore. Besides the skin region, there is a smooth region in the bulk, where density remains low and relatively uniform. 
In addition, the steady-state density $\langle n_A\rangle$ in chain $A$ is precisely the reverse of $\langle n_B\rangle$ in the single-particle case, indicating left skin-localization in chain $A$. When further increasing interchain hopping $\delta$, as depicted in the main plot of Fig. \ref{fig2}(d),  the skin localization gradually weakens. Specifically, the skin region $\zeta$ shrinks and eventually vanishes, while the bulk density correspondingly increases and satisfies $\langle n_{\text{bulk}}\rangle\sim 1/2N$ for relatively large $\delta$, indicating that the steady state has become extended.
For small system size $N$ (e.g. $N=10$), the Liouvillian spectrum and steady state closely resemble the decoupled limit as shown in the Fig. \ref{fig2}(a) and (e), suggesting that coupling effects are negligible. As $N$ increases, the coupling effect emerges. Fig. \ref{fig2}(e) reveals that the skin region of the steady state is almost invariant, while the smooth region gets wider steadily. Furthermore, the purity of the steady state monotonically declines with the increase in $\delta$ (not shown).

Besides the steady-state density distribution, the delocalization process is also reflected in relaxation dynamics.  
With the increase in  $\delta$, Fig. \ref{fig2}(f) illustrates that the Liouvillian gap shifts from being size-independent ($\Delta\propto N^0$) to scaling inversely with the square of the system size ($\Delta\propto 1/N^2$). As discussed in previous works \cite{hagaLiouvillianSkinEffect2021,boundarysensitive}, when the steady state exhibits LSE, the Liouvillian gap is finite in the thermodynamic limit,  and the relaxation time $\tau\propto N$, deviating from the conventional law $\tau\sim 1/\Delta$. Conversely, if the steady state is extended, the Liouvillian gap scales as $\Delta\propto1/N^2$, and the relaxation time satisfies the usual law $\tau\sim 1/\Delta$ again, indicating $\tau\propto N^2$ (diffusive relaxation).

\section{Many-body case}

\begin{figure*}[htb]
\centering
\subfigure{
\includegraphics[height=6.0cm,width=12.0cm]{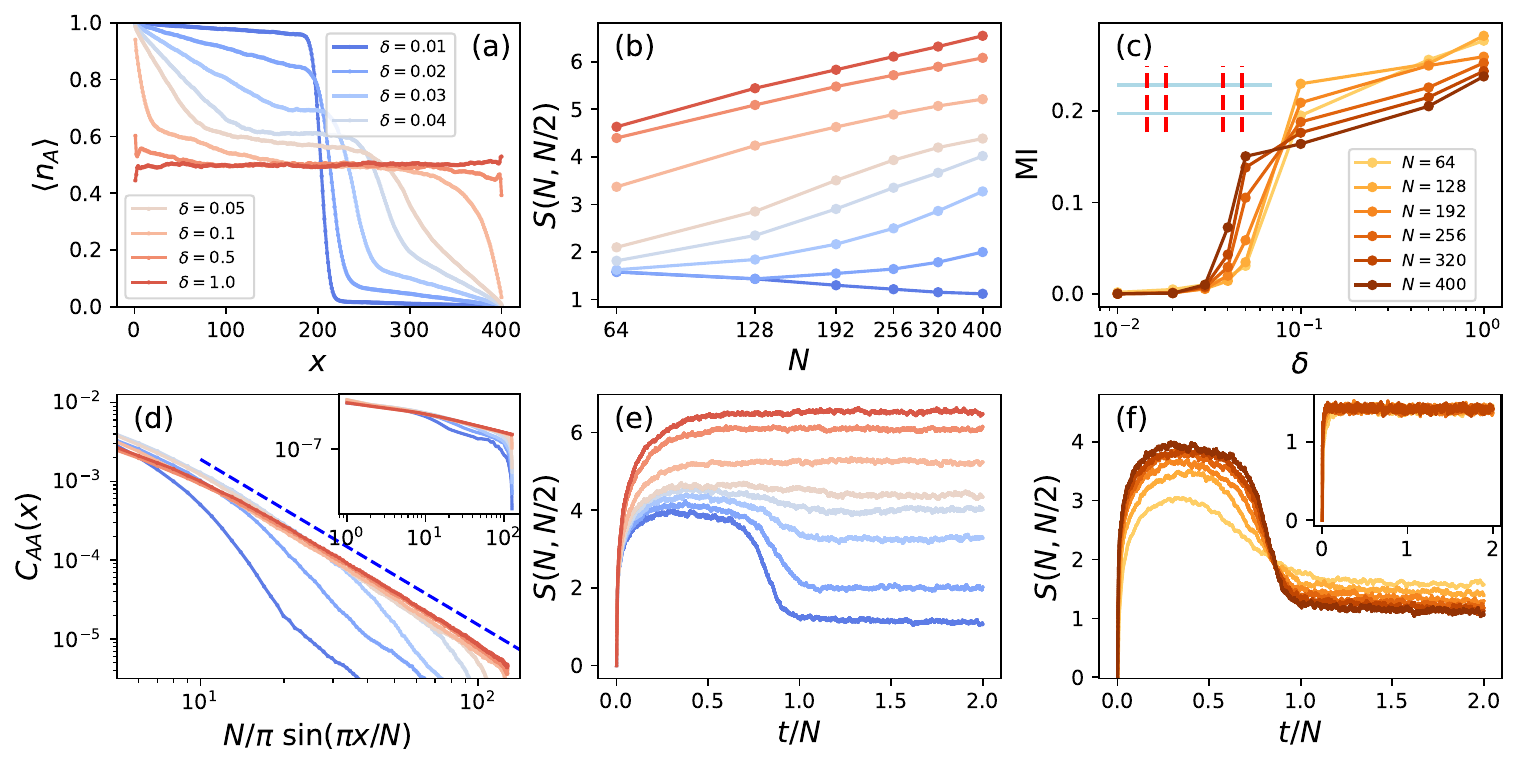}}
\subfigure{
\includegraphics[height=6.0cm,width=4.56cm]{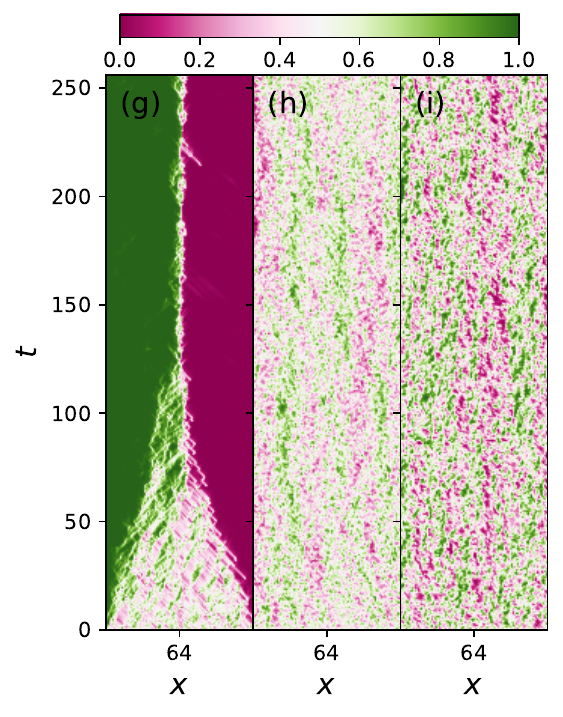}}
\caption{(a) The steady-state density distribution of chain $A$. $N=400$, $\gamma=0.5$. (b) The steady-state entanglement entropy for various system sizes $N$ and interchain couplings $\delta$ in the semi-log plot. $\gamma=0.5$. (c) The steady-state mutual information for various system sizes $N$ and interchain couplings $\delta$ in the semi-log plot. The upper left schematic diagram denotes the two disjoint segments we choose for computing the mutual information. $\gamma=0.5$.
(d) The steady-state connected density-density correlation function in chain $A$ for various $\delta$ in the log-log plot. We truncate the full plot (see the inset) for clarity. The blue dashed line decays in a power-law way and the exponent is about $-2.1$. $\gamma=0.5, N=400$. (e) The evolution of entanglement entropy with time for various $\delta$.  $\gamma=0.5$, $N=400$. (f) The main plot is the evolution of entanglement entropy for various system sizes $N$ when $\gamma=0.5, \delta=0.01$. The inset is the evolution of entanglement entropy for various system sizes $N$ when $\gamma=2.0, \delta=1.0$. (g),(h),(i) A typical trajectory for the evolution of the density of chain $A$ with time. $N=128$ (g) $\gamma=0.5, \delta=0.01$. (h) $\gamma=0.5, \delta=1.0$. (i) $\gamma=2.0, \delta=1.0$. }
\label{fig3}
\end{figure*}

The single-body Lindbladian already demonstrates the delocalization of the skin steady state. In this section, we extend it into the many-body case.
Generally, the whole Lindbladian $\tilde{\mathcal{L}}$ is as large as $(2^{2N})^2\times(2^{2N})^2$,  limiting exact diagonalization to very small sizes in many-particle scenarios. Alternatively, we employ the stochastic Schr\"odinger equation (SSE), which can be regarded as an unraveling of the LME, to approximately simulate the dynamics. In short, SSE rewrites the LME as a stochastic average over many independent trajectories. For each trajectory, the initial pure state evolves according to the non-Hermitian effective Hamiltonian $H_{\text{eff}}$, and the evolution process is interrupted occasionally by quantum jumps modeled by Lindblad operators \cite{MoteCarloWavefunction,quantumtrajectory}. 
The stochastic Schr\"odinger equation is as follows \cite{quantum-measurement-and-control,continuous-quantum-measurement,quantum-noise}:
\begin{equation}\label{SSE}
	\begin{split}
	d|\psi\rangle &= -\mathrm{i} {H}_{\text{eff}}dt|\psi\rangle-\dfrac{\mathrm{i}}{2}dt\langle H_{\text{eff}}^{\dagger}-H_{\text{eff}}\rangle|\psi\rangle + \\
 &\sum_{\mu=\{A,B\}}\sum_{j=1}^{N-1}[\dfrac{{L}_{j,\mu}}{\sqrt{\langle {L}^{\dagger}_{j,\mu}{L}_{j,\mu}\rangle}}-1]|\psi\rangle\,dW_{j,\mu},
		\end{split}
\end{equation}
in which $dW_{j,\mu}$ is a discrete, independent
Poisson random variable demanding $dW_{j,\mu}=0$ or $1$, with mean value $\overline{dW_{j,\mu}}=\gamma\langle{L}_{j,\mu}^{\dagger}{L}_{j,\mu}\rangle dt$.
Given an initial Gaussian state, the SSE (Eq. (\ref{SSE})) preserves the Gaussianity during the evolution as illustrated in Appendix \ref{sec:appendixB}, enabling simulation for sizes up to $2N=800$.
Without loss of the generality, the initial state is chosen as half-filled Neel state $|101010...10\rangle_A|010101...01\rangle_B$. Unless otherwise specified, we still set $\gamma=0.5$ (weak monitoring) henceforth. Additionally, 300 trajectories are simulated for each parameter set ($\{\gamma, \delta, N\}$) to perform the statistical average. To characterize the correlation and entanglement of the system, we calculate the von Neumann entanglement entropy $S(N, N/2)$, von Neumann mutual information (MI), and connected density-density correlation function ($C_{AA}(x)$, $C_{BB}(x)$). The entanglement entropy measures the entanglement between the ladder's left and right halves, while mutual information assesses the correlation between two disjoint segments, each of width $N/8$
and symmetrically positioned 
$N/2$ apart. The connected density-density correlations in chains $A$ and $B$ are defined as $C_{\nu\nu}(x)=\langle n_{N/2,\nu}\rangle\langle n_{N/2+x,\nu}\rangle-\langle n_{N/2,\nu}n_{N/2+x,\nu}\rangle, \nu=\{A, B\}$. The details for the numerical implementation are explained in the Appendix. \ref{sec:appendixB}.

Firstly, we concentrate on the steady state. As shown in Fig. \ref{fig3}(a), for small $\delta$, (e.g. $\delta=0.01$), particles in chain $A$ are predominantly localized on the left side. As $\delta$ increases, the particle density distribution smooths out, becoming nearly uniform for large $\delta$ ($\delta\sim0.5$). Moreover, the steady-state density in chain $B$ undergoes a similar transition as depicted in Fig. \ref{suppfig3}(a).
This indicates that the many-body Liouvillian skin effect diminishes with increasing coupling strength.

As for entanglement entropy $S(N/2, N)$, it has been demonstrated that for $\delta=0$, any nonzero monitoring rate $\gamma$ drives the system into the area law phase due to the suppression effect of LSE on entanglement entropy \cite{MISE,MISE-longrange}. For  $\delta=0.01$, the numerical data suggests that the scaling of entanglement entropy still obeys area law as shown in Fig. \ref{fig3}(b). In contrast, for $\delta=1.0$, the entanglement entropy follows the logarithmic law. For intermediate coupling,  except the frozen region close to the edge caused by LSE, there is a ``fluctuating zone" around the middle (corresponding to the zone in which density distribution is relatively smooth), where particles can move and build correlation \cite{MISE-longrange}. As shown in Fig. \ref{suppfig4}, this ``fluctuating zone" expands with increasing system size $N$, leading to a corresponding growth in entanglement entropy.
Therefore, it is anticipated that the scaling behavior of entanglement will change from area law into logarithmic law by increasing interchain coupling.

The von Neumann mutual information, which provides an upper bound on correlation functions, further supports the delocalization process. 
As depicted in Fig. \ref{fig3}(c), near-zero mutual information for small $\delta$ indicates the negligible correlation between segments due to strong LSE. Increasing $\delta$ reduces the ``frozen effect” of LSE, allowing finite mutual information to emerge, indicating the establishment of correlations between disconnected segments. 

We also examine the connected density-density correlation functions $C_{AA}(x)$ and $C_{BB}(x)$. As shown in Fig. \ref{fig3}(d), $C_{AA}(x)$ exponentially decays for small  $\delta$, but transitions to an algebraic decay ($C_{AA}(x)\propto[\text{sin}(\pi x/N)]^{-2.1}$) when $\delta$ is close to $1.0$. 
These behaviors confirm the transition from the skin steady state to the extended steady state. 
In addition, for small $\delta$, $C_{AA}(x)$ decays abruptly when $x$ is close to $N/2$ (see the inset of Fig. \ref{fig3}(d) and Fig. \ref{suppfig3}(c)), indicating that the particles near the boundary are almost uncorrelated with particles in the bulk owing to the ``frozen effect" of LSE. This sudden drop disappears when the steady state is extended.
Furthermore, for the skin steady state, chain $A$ has fewer particles on the right half compared to chain $B$, rendering correlation harder to establish and causing $C_{AA}(x)$ to decay faster than $C_{BB}(x)$, which is supported in Fig. \ref{suppfig3}(c).
On the contrary, for the extended steady state, there will be no such evident distinction of particle occupation between the right half of chain $A$ and $B$. Consequently, $C_{AA}(x)$ approximately coincides with $C_{BB}(x)$ as shown in the inset of Fig. \ref{suppfig3}(c).

Apart from the steady-state observables, entanglement dynamics also signifies the gradual shift from the skin to the extended steady state. As depicted in Fig. \ref{fig3}(f), for small $\delta$, the entanglement entropy initially grows quickly, indicating that the bulk dynamics dominates and correlations propagate. However, after some time,  the entanglement entropy begins to decrease for a duration, illustrating that the particles touch the boundary and accumulate over time.
With the increase in $\delta$, the entanglement-decreasing process gradually shrinks, which implies that the LSE becomes progressively weaker and finally disappears. Notably,  although the steady state for small $\delta$ follows the area law, the maximum entanglement during the evolution adheres to the logarithmic law \cite{MISE-DPT}, which is the result of weak monitoring.  

Interestingly, our model intriguingly displays two distinct mechanisms for entanglement transition. Apart from the entanglement transition caused by the delocalization of the skin steady state, strong measurements for the extended steady state can induce a conventional MIPT as well. These two different mechanisms are directly reflected in the trajectory. Firstly, as shown in Fig. \ref{fig3}(g), for $\delta=0.01$, $\gamma=0.5$,  particles in chain $A$ accumulate on the left side, indicating that the steady state is skin-localized. Moreover, the entanglement follows area law. As the coupling $\delta$ is increased while maintaining the monitoring rate $\gamma$ constant, the LSE weakens, leading to the change in the scaling behavior of the entanglement. For instance, at $ \delta=1.0, \gamma=0.5$, the steady state is extended [see Fig. \ref{fig3}(h)] and obeys the logarithmic law.
Furthermore, increasing the monitoring rate $\gamma$ while keeping $\delta$ fixed induces another change in the scaling behavior of entanglement from the logarithmic law to the area law, driven by the quantum Zeno effect. For example, as shown in Fig. \ref{fig3}(i), the steady state remains extended, but the local density is closer to $0$ or $1$, compared with weak monitoring in Fig. \ref{fig3}(h). The reason is that strong measurements suppress the unitary evolution imposed by the Hamiltonian and project entangled states to the product states. In other words, measurements project each site into either an occupied or unoccupied state. Conversely, in the limit of weak measurements ($\gamma \rightarrow 0$), unitary evolution benefits entanglement growth. Seldom measurements leave local site occupation indeterminate and the average local density is thus close to $0.5$. 
Moreover, these two different mechanisms are distinguished by entanglement dynamics. The area law induced by LSE experiences the entanglement decreasing process due to particle accumulation at the edges [see the main plot in Fig. \ref{fig3}(f)], whereas the area law induced by the quantum Zeno effect does not show this behavior [see the inset of Fig. \ref{fig3}(f)]. It is also noted that the steady state of our Lindbladian varies with monitoring rate $\gamma$ and coupling strength $\delta$, while the steady state in previous studies on MIPT is always the invariant maximum mixed state $\Bbb{I}$, which implies the generality of the MIPT.

\section{Conclusion and Outlook}
\label{sec:conclsn}

In summary, we investigate the delocalization of skin steady states by proposing an intuitive Lindbladian model comprising two coupled chains with reversed Liouvillian skin localization. Initially, we analyze the single-particle scenario, where the steady-state density distribution and Liouvillian gap obtained via exact diagonalization confirm the gradual transition of the steady state from skin localized to extended. We also uncover the system-size sensitivity of the Liouvillian spectrum, attributable to the system-size dependence of the non-Hermitian effective Hamiltonian. Perturbation analysis predicts a counterpart to the critical non-Hermitian skin effect in open quantum systems, which we term the critical Liouvillian skin effect.
Furthermore, we employ the stochastic Schr\"odinger equation to approximate the dynamics governed by the Lindblad master equation in the many-body half-filled sector. Various indicators, including steady-state density distribution, von Neumann entanglement entropy, mutual information, density-density correlation functions, and entanglement dynamics, all demonstrate the delocalization of the many-body skin steady state. Notably, our model exhibits two distinct mechanisms for inducing entanglement transitions: the conventional measurement-induced phase transition driven by the quantum Zeno effect, and the novel delocalization of 
many-body skin steady state.

In the future, it is tempting to design other simpler and more experimentally feasible setups to investigate this delocalization of skin steady states \cite{hagaLiouvillianSkinEffect2021}. In addition, the impact of the Liouvillian skin effect on quantum criticality and more general non-reciprocal phase transitions remains unclear \cite{QuantumCriticalitynonreciprocity,non-reciprocal-phasetransition}, warranting further investigation. Lastly, given the significant role of dimensionality in the non-Hermitian skin effect \cite{zhangUniversalNonHermitianSkin2022,DimensionalTransmutation,Amoeba,huNonHermitianBandTheory2023,huNHSEhighDimensions}, extending the Liouvillian skin effect to higher dimensions is an intriguing prospect for future study.

\begin{acknowledgments}
We thank Shou Liu and Yupeng Wang for their valuable comments on the manuscript.
The work is supported by National Key Research
and Development Program of China (Grant No.
2021YFA1402104), the NSFC under Grants No.12174436
and No.T2121001 and the Strategic Priority Research
Program of Chinese Academy of Sciences under Grant
No. XDB33000000.
\end{acknowledgments}

%\clearpage
\appendix
\renewcommand{\theequation}{S\arabic{equation}}
\setcounter{equation}{0}
\renewcommand{\thefigure}{S\arabic{figure}}
\setcounter{figure}{0}

\section{Perturbation analysis of the  Lindbladian}
\label{sec:appendixA}

In the main text, it is known that the effective non-Hermitian Hamiltonian is $H_{\text{eff}}=H_{0}-\dfrac{\mathrm{i}}{2}\gamma\sum_{\nu=\{A,B\}}\sum_{j=1}^{N-1}L_{j,\nu}^{\dagger}L_{j,\nu} =\sum_{j=1}^{N-1}(t+\dfrac{\gamma}{4})c_{j,A}^{\dagger}c_{j+1,A}+(t-\dfrac{\gamma}{4})c_{j+1,A}^{\dagger}c_{j,A}
+(t-\dfrac{\gamma}{4})c_{j,B}^{\dagger}c_{j+1,B} +
(t+\dfrac{\gamma}{4})c_{j+1,B}^{\dagger}c_{j,B}
-\dfrac{\mathrm{i}}{4}\gamma(n_{j,A}+n_{j+1,A}+n_{j,B}+n_{j+1,B})+\delta\sum_{j=1}^N (c^{\dagger}_{j,A}c_{j,B}+c^{\dagger}_{j,B}c_{j,A})$. 
We notice that a similar non-Hermitian Hamiltonian $H_{\text{2-chain}}$ has been studied in previous work \cite{CNHSE,scalingCNHSE,auxiliaryGBZ}, which found that arbitrary small coupling between two chains with different NHSE leads to a dramatic change in the energy spectrum and eigenstates in the thermodynamic limit. The phenomenon is called the critical non-Hermitian skin effect (CNHSE), which is explained by the sudden change of the generalized Brillouin Zone \cite{CNHSE,auxiliaryGBZ}. Moreover, the energy spectrum is also sensitive to system size for finite-sized systems \cite{CNHSE,scalingCNHSE}. Interestingly, our effective non-Hermitian Hamiltonian exhibits similar properties as well. 

\begin{figure}[htb]
\includegraphics[width=0.47\textwidth]{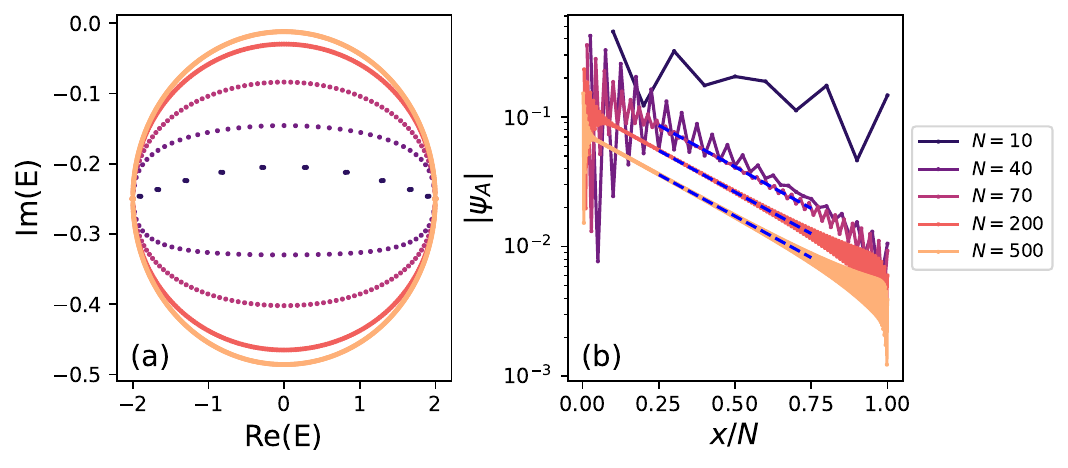}
\caption{$\gamma=0.5, \delta=0.01$. (a) The energy spectrum of non-Hermitian effective Hamiltonian $H_{\text{eff}}$ with different system sizes $N$. (b) The eigenstate of $H_{\text{eff}}$ with the largest Im$(E)$ for various $N$ in the semi-log plot. We only plot the eigenfunction $\psi_A$ in the $A$ chain, while $\psi_B$ is exactly the inverse of $\psi_A$. The fitting blue dashed line is $|\psi_A(x)|\approx0.075e^{-2.95x/N}, 0.115e^{-2.95x/N},  0.18e^{-2.95x/N}$. }
\label{suppfig1}
\end{figure}

As shown in Fig. \ref{suppfig1}(a), the energy spectrum of $H_{\text{eff}}$ varies with system size $N$, with eigenstates corresponding to max[Im($E$)] exhibiting left skin localization in chain $A$ and right skin localization in chain $B$. Furthermore, Fig. \ref{suppfig1}(b) indicates that the skin localization is scale-free, implying $|\psi_A(x)|\propto e^{-\kappa x}$ and $\kappa=\text{const}/N$. In other words, the localization length is proportional to system size $N$.

\begin{figure}[htb]
\includegraphics[width=0.47\textwidth]{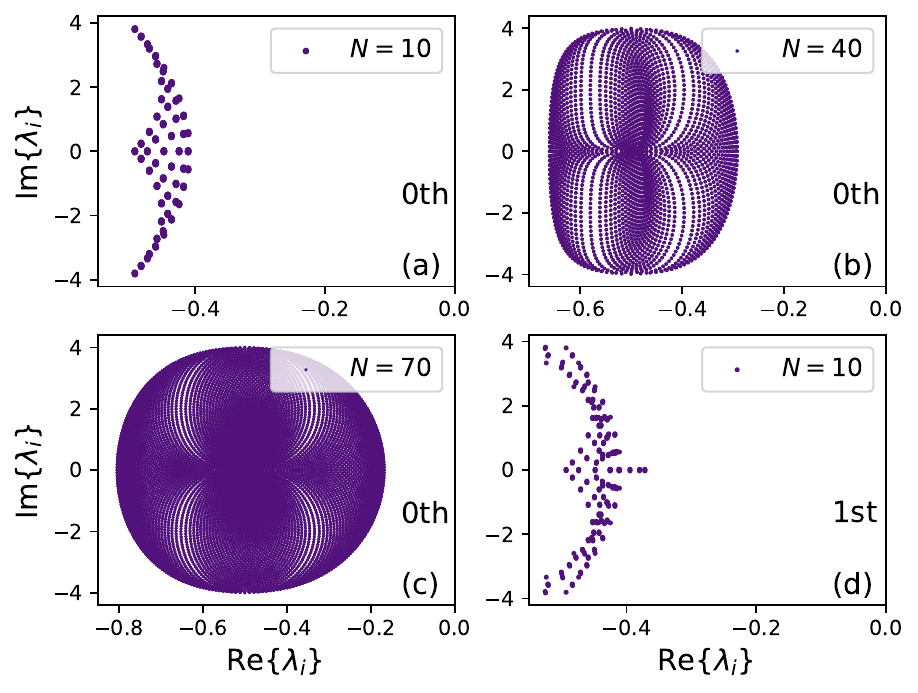}
\caption{$\gamma=0.5, \delta=0.01$. The perturbation analysis of the single-body Lindbladian. (a),(b),(c) The zeroth-order perturbation results. (a) $N=10$. (b) $N=40$. (c) $N=70$. (d) The first-order perturbation results. $N=10$. }
\label{suppfig2}
\end{figure}

This system-size sensitivity of 
$H_{\text{eff}}$ naturally connects to the system-size sensitivity of the Lindbladian discussed in the main text (Fig. \ref{fig2}(a),(b),(c)) by perturbation theory. Using Choi-Jamiolkowski isomorphism \cite{CHOI,JAMIOLKOWSKI}, we vectorize the LME [see Eq. (\ref{LME})] and transform Liouvillian superoperator $\mathcal{L}$ into matrix form $\tilde{\mathcal{L}}$.
\begin{equation}\label{LME2}
	\begin{split}
 \begin{aligned}
&\dfrac{d|\rho\rangle\rangle}{dt}=\tilde{\mathcal{L}}|\rho\rangle\rangle =[-\mathrm{i}(H\otimes I-I\otimes H^{T}) \\
 &+\gamma\sum_{\mu}(L_{j,\nu}\otimes L_{j,\nu}^{*}-\dfrac{1}{2}L_{j,\nu}^{\dagger}L_{j,\nu}\otimes I-\dfrac{1}{2}I\otimes L^{T}_{j,\nu}L^{*}_{j,\nu})]|\rho\rangle\rangle \\
 &=\left[-\mathrm{i}\left(H_{\text{eff}}\otimes I-I\otimes H_{\text{eff}}^* \right)+\gamma\sum_{j,\nu}L_{j,\nu}\otimes L_{j,\nu}^*\right ]|\rho\rangle\rangle,
 \end{aligned}
\end{split}
\end{equation}
 in which $|\rho\rangle\rangle=\sum_{i,j}\rho_{i,j}|i\rangle\otimes|j\rangle$ is the vectorized form of density matrix $\rho=\sum_{i,j}\rho_{i,j}|i\rangle\langle j|$. 
We then decompose the Lindbladian $\tilde{\mathcal{L}}$ into two parts: $\tilde{\mathcal{L}}=\tilde{\mathcal{L}_{0}}+\gamma\tilde{\mathcal{L}_{1}}$. The unperturbed part $\tilde{\mathcal{L}_{0}}$ and perturbed part $\tilde{\mathcal{L}_{1}}$ are defined as follows:
\begin{equation}
\begin{aligned}
\tilde{\mathcal{L}_{0}}&=-\mathrm{i}\left(H_{\text{eff}}\otimes I-I\otimes H_{\text{eff}}^{*} \right) \\
\tilde{\mathcal{L}_{1}}&=\sum_{\nu=\{A,B\}}\sum_{j=1}^{N-1}L_{j,\nu}\otimes L_{j,\nu}^{*}.
\end{aligned}
\end{equation}
The zeroth-order and first-order perturbation results, shown in Fig. \ref{suppfig2}, indicate that the system-size sensitivity of the Lindbladian inherits from the system-size sensitivity of the non-Hermitian effective Hamiltonian.
The zeroth-order perturbation results in Fig. \ref{suppfig2}(a),(b),(c)  indicate that the system-size sensitivity of the Lindbladian inherits from the system-size sensitivity of the non-Hermitian effective Hamiltonian $H_{\text{eff}}$. Furthermore, the first-order perturbation results [see Fig. \ref{suppfig2}(d)] are closer to the exact Liouvillian spectrum [see Fig. \ref{fig2}(a)], compared with the zeroth-order perturbation results [see Fig. \ref{suppfig2}(a)]. However, perturbation results fail to capture the steady state and eigenmodes around the steady state. 

Based on the perturbation theory and critical non-Hermitian skin effect of $H_{\text{eff}}$ \cite{CNHSE}, we anticipate that arbitrary small interchain hopping $\delta$ will cause a dramatic change in the Liouvillian spectrum and steady state in the thermodynamic limit, which we term the critical Liouvillian skin effect. 
The abrupt change of steady state is natural. For zero coupling, the two chains are disconnected. If the particle is in the chain $A$ initially, the particle will be exponentially localized in chain $A$, and chain $B$ remains empty in the steady state. However, any nonzero coupling will connect these two chains, and the steady-state density in chain A and chain B is exactly the opposite.   

At last, we emphasize the significance of studying the skin effect through the Lindbladian rather than the non-Hermitian Hamiltonian. 
Pure non-Hermitian evolution, as used in previous works \cite{EPTNonHermitian,liDisorderInducedEntanglementPhase2023,NHEvoQuasiDisorder,NHEvoAA}, leads to a steady state that is the eigenstate with the maximum imaginary part, i.e., a scale-free exponential localization state as depicted in Fig. \ref{suppfig1}(b).  However, the actual steady state of the Lindbladian [see Fig. \ref{fig2}(d),(e)] is seriously deviated from this, implying that the non-Hermitian evolution based on post-selection fails to grasp the true properties of open quantum systems.

\section{The numerical implementation of quantum jump method. }
\label{sec:appendixB}

In this work, we utilize the stochastic Schr\"odinger equation (SSE) to approximately simulate the dynamics governed by LME [see Eq. (\ref{LME})] in the half-filled sector. The SSE preserves the Gaussianity of the state, as demonstrated below.
A Gaussian state is represented as $|\psi(t)\rangle=\prod_{l=1}^N (\sum_{j=1}^{2N}U_{jl}(t)c_{j}^{\dagger})|0\rangle$, in which $U$ is a $2N\times N$ matrix. In this work, we initialize the state as $|\psi(t=0)\rangle=|1001100110011001...1001\rangle$. We emphasize that we omit the subscript $A, B$ in the main text for convenience ($(n, A) \mapsto 2n-1; (n, B) \mapsto 2n $.)
So the initial states for chains $A$ and $B$, the initial state is $|\psi(t=0)\rangle_{A}=|101010...10\rangle$ and $|\psi(t=0)\rangle_{B}=|010101...01\rangle$, respectively. The evolution is mainly consist of two processes:

(I) Firstly, the system evolves according to the non-Hermitian effective Hamiltonian $H_{\text{eff}}=\sum_{m,n}[h_{\text{eff}}]_{mn}c_m^{\dagger}c_{n}$ for a short time $\delta t$: 
\begin{equation}
\begin{split}
|\psi(t+\delta t)\rangle &= e^{-\mathrm{i}H_\text{eff}\delta t}|\psi(t)\rangle\\
&= \prod\limits_{l=1}^N(\sum^{2N}_{j=1}U_{jl}(t)e^{-\mathrm{i}H_\text{eff}\delta t} c^{\dagger}_j e^{\mathrm{i}H_\text{eff}\delta t})|0\rangle.
\end{split}
\end{equation}
Using the Baker-Campbell-Hausdorff formula, it can be acquired
$e^{-\mathrm{i}\hat{H}_\text{eff}\delta t} c^{\dagger}_j e^{\mathrm{i}\hat{H}_\text{eff}\delta t} = \sum_{m=1}^{2N}\left[ e^{-\mathrm{i}h_\text{eff}\delta t}\right]_{mj}c^{\dagger}_m$. Therefore, the above equation is reduced to: 
\begin{equation}
\begin{split}
&= \prod\limits_{l=1}^N(\sum^{2N}_{j=1}U_{jl}(t)\sum_{m=1}^{2N} \left[e^{-ih_\text{eff}\delta t}\right]_{mj}c^{\dagger}_m)|0\rangle\\
&=\prod\limits_{l=1}^N\sum_{m=1}^{2N}\left[e^{-ih_\text{eff}\delta t}U\right]_{ml}c^{\dagger}_m|0\rangle .
\end{split}
\end{equation}
Hence, $U(t+\delta t) = e^{-\mathrm{i}h_\text{eff}\delta t}U(t)$. To preserve $U^{\dagger} U = I$, we can perform a $QR$ decomposition, $U(t+\delta t) = \text{QR}$, and reassign $U(t+\delta t)$ as $Q$.

(II) Next, consider the actions of quantum jumps. There are $2(N-1)$ Lindblad operators $\{L_{1}, L_{2}, L_{3}, L_{4},...,L_{2N-2}\}$, corresponding to $\{L_{1,A},L_{1,B},L_{2,A},L_{2,B}, ...L_{N-1,A},L_{N-1,B}\}$ in the main text. The Lindblad operators act as follows: 
\begin{equation}
	\begin{split}
	U(t+\delta t)\rangle = M_{\delta t}[e^{-\mathrm{i}H_\text{eff}\delta t}|U(t)\rangle],\\
	M_{\delta t}[|U\rangle] = \prod\limits_{i\in P}\dfrac{L_i|U\rangle}{\parallel L_i|U\rangle\parallel},
		\end{split}
	\end{equation}
where $P=\{n|r_n<\gamma\langle L^{\dagger}_nL_n\rangle\delta t\}$, $r_n \in (0,1)$ is a set of independent random variables.
In detail, after the non-Hermitian evolution in step (I), we generate a set of random numbers $r_n$ to decide whether quantum jump ${L}_n$ occurs.
Assuming ${L}_i=e^{\mathrm{i}\pi{n}_{i+1}} {\xi}_{i}^{\dagger}{\xi}_{i}$ and ${\xi}^{\dagger}_i=\sum_k[a_i]_{k}{c}_{k}^{\dagger}$ (here, we neglect the subscripts of $A$ and $B$ for convenience),
the occurring probability of the quantum jump ${L}_{i}$ is
$p_i = \langle\psi|{L}_i^{\dagger}{L}_i|\psi\rangle\gamma\delta t$ = $\langle\psi|{\xi}^{\dagger}_i{\xi}_i|\psi\rangle\gamma\delta t$. We find:  
\begin{equation}
\begin{split}
	{\xi}_i|\psi\rangle = (\sum_k[a_i]^*_{k}\hat{c}_k)\prod_{l=1}^N(\sum_{j=1}^{2N}U_{jl}(t)c^{\dagger}_j)|0\rangle\\
	=\sum_k\langle a_i|U_k\rangle\bigotimes_{j\neq i}|U_j\rangle.
\end{split}
\end{equation}
Because of $U^{\dagger}U = I$, we obtain $p_i=\gamma\delta t\sum_{k}|\langle a_i|U_{k}\rangle|^{2}$.
To simplify the expression, we recall that the state is invariant for the elementary column
operations. So we can pre-orthogonalize: find the first column $k$ that satisfies $\langle a_i|U_k\rangle\neq 0$, then move the column $k$ into the first column, and transform the other columns as:
\begin{equation}
	\begin{split}
	|U_{j}^{\prime}\rangle &= |U_j\rangle-\dfrac{\langle a_i|U_j\rangle}{\langle a_i|U_1\rangle}|U_1\rangle,
		\end{split}
	\end{equation}
Therefore, $\forall j\geq 2$, \, $\langle a_i|U_{j}^{'} \rangle = 0$, and ${\xi}_{i}|\psi\rangle= \bigotimes_{i\geq 2}|U_{i}^{'}\rangle$. After applying the Lindblad operator, we get ${L}_{i}|\psi\rangle=e^{\mathrm{i}\pi{n}_{i+1}}{\xi}_{i}^{\dagger}{\xi}_{i}|\psi\rangle=e^{\mathrm{i}\pi {n}_{i+1}}\left[| a_i\rangle\bigotimes_{j\geq 2}|U_{j}^{'}\rangle\right]=|e^{\mathrm{i}\pi M} a_i\rangle\bigotimes_{j\geq 2}|e^{\mathrm{i}\pi M}U_{j}^{\prime}\rangle$, in which $M$ is a $2N \times 2N$ matrix representing $n_{i+1}$. After the quantum jump ${L}_{i}$, we perform the $QR$ decomposition again and reassign $U$ as $Q$.

These processes (I) and (II) are repeated until the system reaches a steady state. In addition, we compute four observables: density distribution, von Neumann
entanglement entropy $S(N, N/2)$, von Neumann mutual
information (MI) and connected density-density correlations, all of which can be acquired from the two-point correlation function.
The two-point correlation function $D_{ij}(t)=\langle\psi(t)|c^{\dagger}_ic_j|\psi(t)\rangle=[U(t)U^{\dagger}(t)]_{ji}, i,j=1,2,3,...2N$ allows us to extract the local density from its diagonal elements. The von Neumann entanglement entropy $S(N, N/2)$ is computed from the eigenvalues of the subsystem's two-point correlation function $D_{\text{sub}}$. We choose the subsystem as the left-half side of the ladder so that the entanglement entropy is expressed as: 
\begin{equation}
S(N, N/2)=-\sum_{k=1}^N \eta_k\text{log}\eta_k+(1-\eta_k)\text{log}(1-\eta_k),
\end{equation}
in which $\{\eta_k\}$ are the eigenvalues of $N\times N$  two-point correlation function of subsystem $[D]_{i\in[1,N],j\in[1,N] }$.
As for von Neumann mutual information, we select two disjoint segments (denoted as region V and VI) as depicted in the schematic diagram of Fig. \ref{fig3}(c). The two segments are $N/8$ wide and they are symmetrically positioned with a separation distance of $N/2$. The mutual information is defined as $\text{MI}=S(\text{V})+S(\text{VI})-S(\text{V}\cup \text{VI})$, in which $S(\text{V})$ represents the entanglement between the region V and its complement. The entanglement entropy $S(\text{V}), S(\text{VI}), S(\text{V}\cup \text{VI})$ are computed in the same manner as $S(N, N/2)$. Moreover, using Wick theorem, the connected density-density correlation function is reduced to  
$C_{\nu\nu}(x)=\langle n_{N/2,\nu}\rangle\langle n_{N/2+x,\nu}\rangle-\langle n_{N/2,\nu}n_{N/2+x,\nu}\rangle=|\langle c_{N/2,\nu}^{\dagger}c_{N/2+x,\nu}\rangle|^2, \nu=\{A, B\}$. 

In this work, we set the time step $\delta t=0.05$ and total evolution time $t_{\text{tot}}=2N$, which is sufficiently long to ensure the system reaches a steady state. The primary source of numerical error arises from the discrete-time step $\delta t$; hence, we compare the results for $\delta t=0.01$ and $\delta t=0.05$, observing no qualitative differences. We also use several initial states to verify the uniqueness of the steady state.
For each set of parameters ($\gamma, \delta, N$), we calculate $300$ trajectories to reduce statistical errors. 

\begin{figure}[htb]
\includegraphics[width=0.47\textwidth]{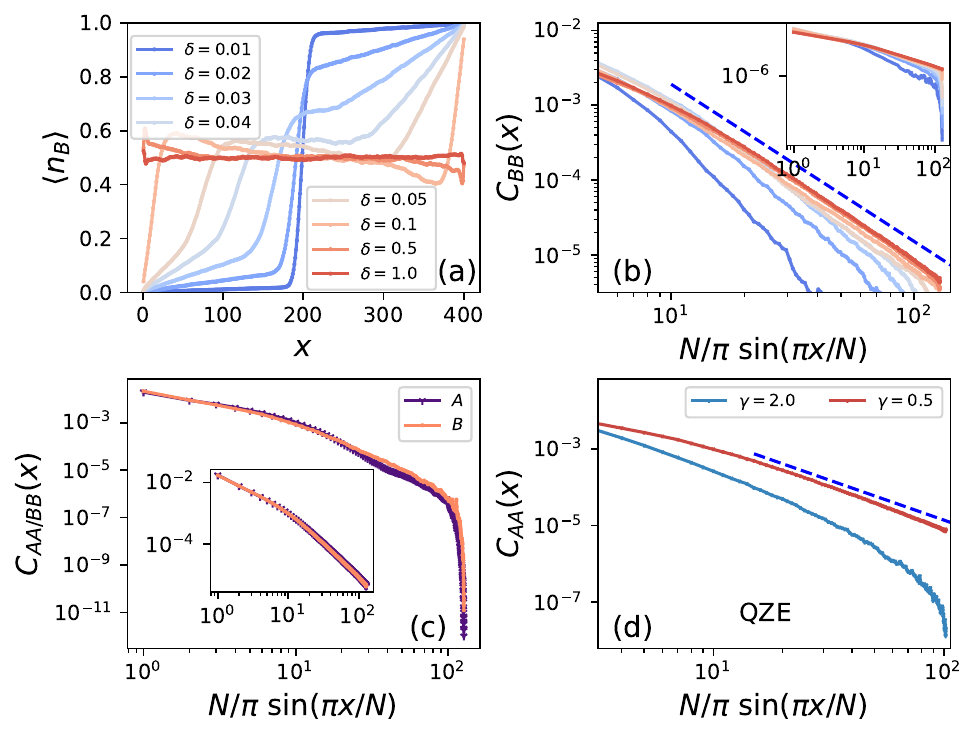}
\caption{(a) The steady-state density distribution of chain $B$ for various interchain coupling $\delta$. $\gamma=0.5, N=400$. (b) The steady-state density-density correlation function of chain $B$ for various $\delta$ in a log-log plot. $\gamma=0.5, N=400$. The blue dashed line decays in a power-law way and the exponent is about $- 2.1$. (c) The comparison of the $C_{AA}(x)$ and $C_{BB}(x)$ in the skin localized phase (main plot $\delta=0.02$) and the extended phase (inset $\delta=1.0$) in a log-log plot. $\gamma=0.05, N=400$. (d) $N=320,  \delta=1.0$. In the extended phase, for weak monitoring $\gamma=0.5$, the correlation function decays in a power-law way, while for strong measurement $\gamma=2.0$, the correlation function exponentially decays. }
\label{suppfig3}
\end{figure}

\section{More supporting numerical results}\label{sec:appendixC}

\begin{figure}[htb]
\centering
\includegraphics[width=0.47\textwidth]{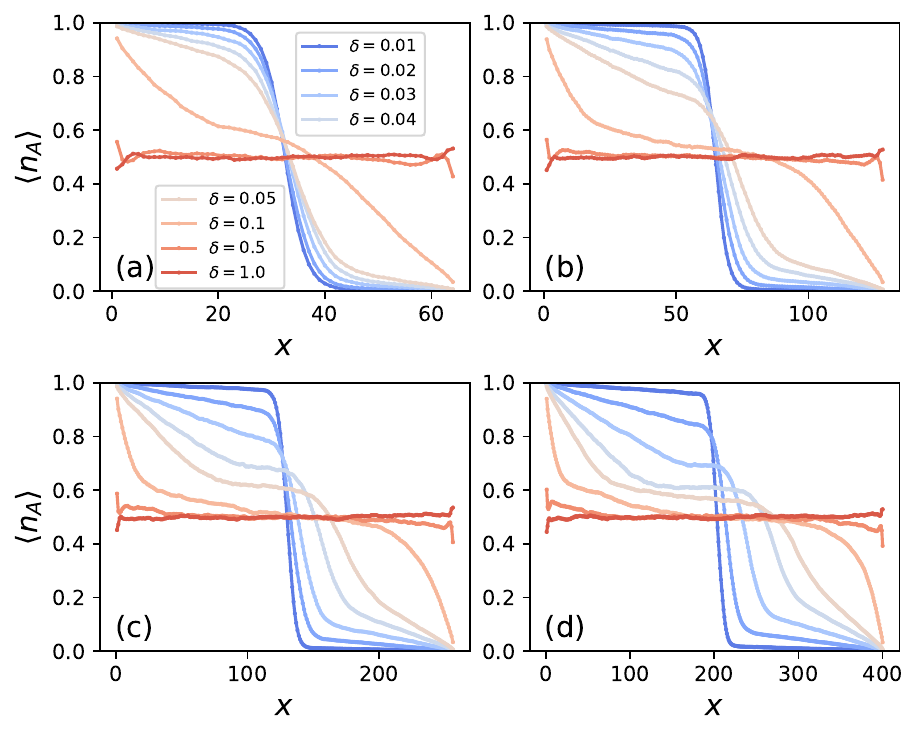}
\caption{The steady-state density distribution of chain $A$. $\gamma=0.5, \delta=0.01$. (a) $N=64$. (b) $N=128$. (c) $N=256$. (d) $N=400$. }
\label{suppfig4}
\end{figure}

The main text presents the steady-state density distribution $\langle n_A\rangle$ and density-density correlation function $C_{AA}(x)$.
To further substantiate the delocalization of skin steady state, we provide additional numerical results for chain $B$ ($\langle n_B\rangle$, $C_{BB}(x)$). As shown in Fig. \ref{suppfig3} (a), the steady-state density distribution $\langle n_B\rangle$ changes from the right skin-localized into extended as the coupling strength increases ($\langle n_{x,A}\rangle\neq\langle n_{N+1-x,B}\rangle$ for many-body case). Analogous to $C_{AA}(x)$, the $C_{BB}(x)$ in Fig. \ref{suppfig3}(b) changes from exponential decay into algebraic decay. Moreover, as shown in Fig. \ref{suppfig3}(c), $C_{AA}(x)$ decays faster than $C_{BB}(x)$ for the skin steady state, while $C_{AA}(x)$ approximately coincides with $C_{BB}(x)$ for the extended steady state. Furthermore, the entanglement transition induced by the quantum Zeno effect is also signified by the transition of correlation function from exponential to algebraic decay as depicted in Fig. \ref{suppfig3}(d).

To provide insight into the system-size dependence of the many-body Liouvillian skin effect, we plot the steady-state density distribution $\langle n_A\rangle$ for various system sizes $N$ in Fig. \ref{suppfig4}. For medium-strength coupling, e.g. $\delta=0.04$, the relatively smooth region around the middle widens as the system size $N$ grows, indicating that more particles can move and build correlations with each other. Similar behavior is observed for $\langle n_B\rangle$ (not shown). 
Consequently, as shown in Fig. \ref{fig3}(b), the entanglement also increases with $N$ for medium-strength coupling strength.

\bibliographystyle{apsreve}
\bibliography{ref}

\begin{thebibliography}{58}%
\makeatletter
\providecommand \@ifxundefined [1]{%
 \@ifx{#1\undefined}
}%
\providecommand \@ifnum [1]{%
 \ifnum #1\expandafter \@firstoftwo
 \else \expandafter \@secondoftwo
 \fi
}%
\providecommand \@ifx [1]{%
 \ifx #1\expandafter \@firstoftwo
 \else \expandafter \@secondoftwo
 \fi
}%
\providecommand \natexlab [1]{#1}%
\providecommand \enquote  [1]{``#1''}%
\providecommand \bibnamefont  [1]{#1}%
\providecommand \bibfnamefont [1]{#1}%
\providecommand \citenamefont [1]{#1}%
\providecommand \href@noop [0]{\@secondoftwo}%
\providecommand \href [0]{\begingroup \@sanitize@url \@href}%
\providecommand \@href[1]{\@@startlink{#1}\@@href}%
\providecommand \@@href[1]{\endgroup#1\@@endlink}%
\providecommand \@sanitize@url [0]{\catcode `\\12\catcode `\$12\catcode
  `\&12\catcode `\#12\catcode `\^12\catcode `\_12\catcode `\%12\relax}%
\providecommand \@@startlink[1]{}%
\providecommand \@@endlink[0]{}%
\providecommand \url  [0]{\begingroup\@sanitize@url \@url }%
\providecommand \@url [1]{\endgroup\@href {#1}{\urlprefix }}%
\providecommand \urlprefix  [0]{URL }%
\providecommand \Eprint [0]{\href }%
\providecommand \doibase [0]{https://doi.org/}%
\providecommand \selectlanguage [0]{\@gobble}%
\providecommand \bibinfo  [0]{\@secondoftwo}%
\providecommand \bibfield  [0]{\@secondoftwo}%
\providecommand \translation [1]{[#1]}%
\providecommand \BibitemOpen [0]{}%
\providecommand \bibitemStop [0]{}%
\providecommand \bibitemNoStop [0]{.\EOS\space}%
\providecommand \EOS [0]{\spacefactor3000\relax}%
\providecommand \BibitemShut  [1]{\csname bibitem#1\endcsname}%
\let\auto@bib@innerbib\@empty
%</preamble>
\bibitem [{\citenamefont {Ashida}\ \emph {et~al.}(2020)\citenamefont {Ashida},
  \citenamefont {Gong},\ and\ \citenamefont {Ueda}}]{ashida2020non}%
  \BibitemOpen
  \bibfield  {author} {\bibinfo {author} {\bibfnamefont {Y.}~\bibnamefont
  {Ashida}}, \bibinfo {author} {\bibfnamefont {Z.}~\bibnamefont {Gong}},\ and\
  \bibinfo {author} {\bibfnamefont {M.}~\bibnamefont {Ueda}},\ }\bibfield
  {title} {\bibinfo {title} {Non-hermitian physics},\ }\href
  {https://doi.org/10.1080/00018732.2021.1876991} {\bibfield  {journal}
  {\bibinfo  {journal} {Advances in Physics}\ }\textbf {\bibinfo {volume}
  {69}},\ \bibinfo {pages} {249} (\bibinfo {year} {2020})}\BibitemShut
  {NoStop}%
\bibitem [{\citenamefont {Yao}\ and\ \citenamefont {Wang}(2018)}]{GBZ}%
  \BibitemOpen
  \bibfield  {author} {\bibinfo {author} {\bibfnamefont {S.}~\bibnamefont
  {Yao}}\ and\ \bibinfo {author} {\bibfnamefont {Z.}~\bibnamefont {Wang}},\
  }\bibfield  {title} {\bibinfo {title} {Edge states and topological invariants
  of non-hermitian systems},\ }\href
  {https://doi.org/10.1103/PhysRevLett.121.086803} {\bibfield  {journal}
  {\bibinfo  {journal} {Phys. Rev. Lett.}\ }\textbf {\bibinfo {volume} {121}},\
  \bibinfo {pages} {086803} (\bibinfo {year} {2018})}\BibitemShut {NoStop}%
\bibitem [{\citenamefont {Guo}\ \emph {et~al.}(2021)\citenamefont {Guo},
  \citenamefont {Liu}, \citenamefont {Zhao}, \citenamefont {Liu},\ and\
  \citenamefont {Chen}}]{boundarysensitive}%
  \BibitemOpen
  \bibfield  {author} {\bibinfo {author} {\bibfnamefont {C.-X.}\ \bibnamefont
  {Guo}}, \bibinfo {author} {\bibfnamefont {C.-H.}\ \bibnamefont {Liu}},
  \bibinfo {author} {\bibfnamefont {X.-M.}\ \bibnamefont {Zhao}}, \bibinfo
  {author} {\bibfnamefont {Y.}~\bibnamefont {Liu}},\ and\ \bibinfo {author}
  {\bibfnamefont {S.}~\bibnamefont {Chen}},\ }\bibfield  {title} {\bibinfo
  {title} {Exact solution of non-hermitian systems with generalized boundary
  conditions: Size-dependent boundary effect and fragility of the skin
  effect},\ }\href {https://doi.org/10.1103/PhysRevLett.127.116801} {\bibfield
  {journal} {\bibinfo  {journal} {Phys. Rev. Lett.}\ }\textbf {\bibinfo
  {volume} {127}},\ \bibinfo {pages} {116801} (\bibinfo {year}
  {2021})}\BibitemShut {NoStop}%
\bibitem [{\citenamefont {Zhang}\ \emph {et~al.}(2020)\citenamefont {Zhang},
  \citenamefont {Yang},\ and\ \citenamefont {Fang}}]{WindingSkin}%
  \BibitemOpen
  \bibfield  {author} {\bibinfo {author} {\bibfnamefont {K.}~\bibnamefont
  {Zhang}}, \bibinfo {author} {\bibfnamefont {Z.}~\bibnamefont {Yang}},\ and\
  \bibinfo {author} {\bibfnamefont {C.}~\bibnamefont {Fang}},\ }\bibfield
  {title} {\bibinfo {title} {Correspondence between winding numbers and skin
  modes in non-hermitian systems},\ }\href
  {https://doi.org/10.1103/PhysRevLett.125.126402} {\bibfield  {journal}
  {\bibinfo  {journal} {Phys. Rev. Lett.}\ }\textbf {\bibinfo {volume} {125}},\
  \bibinfo {pages} {126402} (\bibinfo {year} {2020})}\BibitemShut {NoStop}%
\bibitem [{\citenamefont {Okuma}\ \emph {et~al.}(2020)\citenamefont {Okuma},
  \citenamefont {Kawabata}, \citenamefont {Shiozaki},\ and\ \citenamefont
  {Sato}}]{windingSkin2}%
  \BibitemOpen
  \bibfield  {author} {\bibinfo {author} {\bibfnamefont {N.}~\bibnamefont
  {Okuma}}, \bibinfo {author} {\bibfnamefont {K.}~\bibnamefont {Kawabata}},
  \bibinfo {author} {\bibfnamefont {K.}~\bibnamefont {Shiozaki}},\ and\
  \bibinfo {author} {\bibfnamefont {M.}~\bibnamefont {Sato}},\ }\bibfield
  {title} {\bibinfo {title} {Topological origin of non-hermitian skin
  effects},\ }\href {https://doi.org/10.1103/PhysRevLett.124.086801} {\bibfield
   {journal} {\bibinfo  {journal} {Phys. Rev. Lett.}\ }\textbf {\bibinfo
  {volume} {124}},\ \bibinfo {pages} {086801} (\bibinfo {year}
  {2020})}\BibitemShut {NoStop}%
\bibitem [{\citenamefont {Borgnia}\ \emph {et~al.}(2020)\citenamefont
  {Borgnia}, \citenamefont {Kruchkov},\ and\ \citenamefont {Slager}}]{Slager}%
  \BibitemOpen
  \bibfield  {author} {\bibinfo {author} {\bibfnamefont {D.~S.}\ \bibnamefont
  {Borgnia}}, \bibinfo {author} {\bibfnamefont {A.~J.}\ \bibnamefont
  {Kruchkov}},\ and\ \bibinfo {author} {\bibfnamefont {R.-J.}\ \bibnamefont
  {Slager}},\ }\bibfield  {title} {\bibinfo {title} {Non-hermitian boundary
  modes and topology},\ }\href {https://doi.org/10.1103/PhysRevLett.124.056802}
  {\bibfield  {journal} {\bibinfo  {journal} {Phys. Rev. Lett.}\ }\textbf
  {\bibinfo {volume} {124}},\ \bibinfo {pages} {056802} (\bibinfo {year}
  {2020})}\BibitemShut {NoStop}%
\bibitem [{\citenamefont {Yokomizo}\ and\ \citenamefont
  {Murakami}(2019)}]{non-bloch}%
  \BibitemOpen
  \bibfield  {author} {\bibinfo {author} {\bibfnamefont {K.}~\bibnamefont
  {Yokomizo}}\ and\ \bibinfo {author} {\bibfnamefont {S.}~\bibnamefont
  {Murakami}},\ }\bibfield  {title} {\bibinfo {title} {Non-bloch band theory of
  non-hermitian systems},\ }\href
  {https://doi.org/10.1103/PhysRevLett.123.066404} {\bibfield  {journal}
  {\bibinfo  {journal} {Phys. Rev. Lett.}\ }\textbf {\bibinfo {volume} {123}},\
  \bibinfo {pages} {066404} (\bibinfo {year} {2019})}\BibitemShut {NoStop}%
\bibitem [{\citenamefont {Yang}\ \emph {et~al.}(2020)\citenamefont {Yang},
  \citenamefont {Zhang}, \citenamefont {Fang},\ and\ \citenamefont
  {Hu}}]{auxiliaryGBZ}%
  \BibitemOpen
  \bibfield  {author} {\bibinfo {author} {\bibfnamefont {Z.}~\bibnamefont
  {Yang}}, \bibinfo {author} {\bibfnamefont {K.}~\bibnamefont {Zhang}},
  \bibinfo {author} {\bibfnamefont {C.}~\bibnamefont {Fang}},\ and\ \bibinfo
  {author} {\bibfnamefont {J.}~\bibnamefont {Hu}},\ }\bibfield  {title}
  {\bibinfo {title} {Non-hermitian bulk-boundary correspondence and auxiliary
  generalized brillouin zone theory},\ }\href
  {https://doi.org/10.1103/PhysRevLett.125.226402} {\bibfield  {journal}
  {\bibinfo  {journal} {Phys. Rev. Lett.}\ }\textbf {\bibinfo {volume} {125}},\
  \bibinfo {pages} {226402} (\bibinfo {year} {2020})}\BibitemShut {NoStop}%
\bibitem [{\citenamefont {Zhang}\ \emph {et~al.}(2022)\citenamefont {Zhang},
  \citenamefont {Yang},\ and\ \citenamefont
  {Fang}}]{zhangUniversalNonHermitianSkin2022}%
  \BibitemOpen
  \bibfield  {author} {\bibinfo {author} {\bibfnamefont {K.}~\bibnamefont
  {Zhang}}, \bibinfo {author} {\bibfnamefont {Z.}~\bibnamefont {Yang}},\ and\
  \bibinfo {author} {\bibfnamefont {C.}~\bibnamefont {Fang}},\ }\bibfield
  {title} {\bibinfo {title} {Universal non-hermitian skin effect in two and
  higher dimensions},\ }\href {https://doi.org/10.1038/s41467-022-30161-6}
  {\bibfield  {journal} {\bibinfo  {journal} {Nature Communications}\ }\textbf
  {\bibinfo {volume} {13}},\ \bibinfo {pages} {2496} (\bibinfo {year}
  {2022})}\BibitemShut {NoStop}%
\bibitem [{\citenamefont {Jiang}\ and\ \citenamefont
  {Lee}(2023)}]{DimensionalTransmutation}%
  \BibitemOpen
  \bibfield  {author} {\bibinfo {author} {\bibfnamefont {H.}~\bibnamefont
  {Jiang}}\ and\ \bibinfo {author} {\bibfnamefont {C.~H.}\ \bibnamefont
  {Lee}},\ }\bibfield  {title} {\bibinfo {title} {Dimensional transmutation
  from non-hermiticity},\ }\href
  {https://doi.org/10.1103/PhysRevLett.131.076401} {\bibfield  {journal}
  {\bibinfo  {journal} {Phys. Rev. Lett.}\ }\textbf {\bibinfo {volume} {131}},\
  \bibinfo {pages} {076401} (\bibinfo {year} {2023})}\BibitemShut {NoStop}%
\bibitem [{\citenamefont {Wang}\ \emph
  {et~al.}(2024{\natexlab{a}})\citenamefont {Wang}, \citenamefont {Song},\ and\
  \citenamefont {Wang}}]{Amoeba}%
  \BibitemOpen
  \bibfield  {author} {\bibinfo {author} {\bibfnamefont {H.-Y.}\ \bibnamefont
  {Wang}}, \bibinfo {author} {\bibfnamefont {F.}~\bibnamefont {Song}},\ and\
  \bibinfo {author} {\bibfnamefont {Z.}~\bibnamefont {Wang}},\ }\bibfield
  {title} {\bibinfo {title} {Amoeba formulation of non-bloch band theory in
  arbitrary dimensions},\ }\href {https://doi.org/10.1103/PhysRevX.14.021011}
  {\bibfield  {journal} {\bibinfo  {journal} {Phys. Rev. X}\ }\textbf {\bibinfo
  {volume} {14}},\ \bibinfo {pages} {021011} (\bibinfo {year}
  {2024}{\natexlab{a}})}\BibitemShut {NoStop}%
\bibitem [{\citenamefont {Hu}(2023)}]{huNonHermitianBandTheory2023}%
  \BibitemOpen
  \bibfield  {author} {\bibinfo {author} {\bibfnamefont {H.}~\bibnamefont
  {Hu}},\ }\bibfield  {title} {\bibinfo {title} {Non-hermitian band theory in
  all dimensions: uniform spectra and skin effect},\ }\href
  {https://arxiv.org/abs/2306.12022} {\bibfield  {journal} {\bibinfo  {journal}
  {arXiv:2306.12022}\ } (\bibinfo {year} {2023})}\BibitemShut {NoStop}%
\bibitem [{\citenamefont {Xiong}\ \emph {et~al.}(2024)\citenamefont {Xiong},
  \citenamefont {Xing},\ and\ \citenamefont {Hu}}]{huNHSEhighDimensions}%
  \BibitemOpen
  \bibfield  {author} {\bibinfo {author} {\bibfnamefont {Y.}~\bibnamefont
  {Xiong}}, \bibinfo {author} {\bibfnamefont {Z.-Y.}\ \bibnamefont {Xing}},\
  and\ \bibinfo {author} {\bibfnamefont {H.}~\bibnamefont {Hu}},\ }\bibfield
  {title} {\bibinfo {title} {Non-hermitian skin effect in arbitrary dimensions:
  non-bloch band theory and classification},\ }\href
  {https://arxiv.org/abs/2407.01296} {\bibfield  {journal} {\bibinfo  {journal}
  {arXiv:2407.01296}\ } (\bibinfo {year} {2024})}\BibitemShut {NoStop}%
\bibitem [{\citenamefont {Dalibard}\ \emph {et~al.}(1992)\citenamefont
  {Dalibard}, \citenamefont {Castin},\ and\ \citenamefont
  {M\o{}lmer}}]{MoteCarloWavefunction}%
  \BibitemOpen
  \bibfield  {author} {\bibinfo {author} {\bibfnamefont {J.}~\bibnamefont
  {Dalibard}}, \bibinfo {author} {\bibfnamefont {Y.}~\bibnamefont {Castin}},\
  and\ \bibinfo {author} {\bibfnamefont {K.}~\bibnamefont {M\o{}lmer}},\
  }\bibfield  {title} {\bibinfo {title} {Wave-function approach to dissipative
  processes in quantum optics},\ }\href
  {https://doi.org/10.1103/PhysRevLett.68.580} {\bibfield  {journal} {\bibinfo
  {journal} {Phys. Rev. Lett.}\ }\textbf {\bibinfo {volume} {68}},\ \bibinfo
  {pages} {580} (\bibinfo {year} {1992})}\BibitemShut {NoStop}%
\bibitem [{\citenamefont {Daley}(2014)}]{quantumtrajectory}%
  \BibitemOpen
  \bibfield  {author} {\bibinfo {author} {\bibfnamefont {A.~J.}\ \bibnamefont
  {Daley}},\ }\bibfield  {title} {\bibinfo {title} {Quantum trajectories and
  open many-body quantum systems},\ }\href
  {https://doi.org/10.1080/00018732.2014.933502} {\bibfield  {journal}
  {\bibinfo  {journal} {Advances in Physics}\ }\textbf {\bibinfo {volume}
  {63}},\ \bibinfo {pages} {77} (\bibinfo {year} {2014})}\BibitemShut {NoStop}%
\bibitem [{\citenamefont {McDonald}\ \emph {et~al.}(2022)\citenamefont
  {McDonald}, \citenamefont {Hanai},\ and\ \citenamefont
  {Clerk}}]{LSEFermionBoson}%
  \BibitemOpen
  \bibfield  {author} {\bibinfo {author} {\bibfnamefont {A.}~\bibnamefont
  {McDonald}}, \bibinfo {author} {\bibfnamefont {R.}~\bibnamefont {Hanai}},\
  and\ \bibinfo {author} {\bibfnamefont {A.~A.}\ \bibnamefont {Clerk}},\
  }\bibfield  {title} {\bibinfo {title} {Nonequilibrium stationary states of
  quantum non-hermitian lattice models},\ }\href
  {https://doi.org/10.1103/PhysRevB.105.064302} {\bibfield  {journal} {\bibinfo
   {journal} {Phys. Rev. B}\ }\textbf {\bibinfo {volume} {105}},\ \bibinfo
  {pages} {064302} (\bibinfo {year} {2022})}\BibitemShut {NoStop}%
\bibitem [{\citenamefont {Ehrhardt}\ and\ \citenamefont
  {Larson}(2024)}]{NHfluctuation}%
  \BibitemOpen
  \bibfield  {author} {\bibinfo {author} {\bibfnamefont {C.}~\bibnamefont
  {Ehrhardt}}\ and\ \bibinfo {author} {\bibfnamefont {J.}~\bibnamefont
  {Larson}},\ }\bibfield  {title} {\bibinfo {title} {Exploring the impact of
  fluctuation-induced criticality on non-hermitian skin effect and quantum
  sensors},\ }\href {https://doi.org/10.1103/PhysRevResearch.6.023135}
  {\bibfield  {journal} {\bibinfo  {journal} {Phys. Rev. Res.}\ }\textbf
  {\bibinfo {volume} {6}},\ \bibinfo {pages} {023135} (\bibinfo {year}
  {2024})}\BibitemShut {NoStop}%
\bibitem [{\citenamefont {Feng}\ and\ \citenamefont
  {Chen}(2024)}]{BoundarySensitiveLindb}%
  \BibitemOpen
  \bibfield  {author} {\bibinfo {author} {\bibfnamefont {X.}~\bibnamefont
  {Feng}}\ and\ \bibinfo {author} {\bibfnamefont {S.}~\bibnamefont {Chen}},\
  }\bibfield  {title} {\bibinfo {title} {Boundary-sensitive lindbladians and
  relaxation dynamics},\ }\href {https://doi.org/10.1103/PhysRevB.109.014313}
  {\bibfield  {journal} {\bibinfo  {journal} {Phys. Rev. B}\ }\textbf {\bibinfo
  {volume} {109}},\ \bibinfo {pages} {014313} (\bibinfo {year}
  {2024})}\BibitemShut {NoStop}%
\bibitem [{\citenamefont {Li}\ \emph {et~al.}(2023{\natexlab{a}})\citenamefont
  {Li}, \citenamefont {Wu}, \citenamefont {Zheng},\ and\ \citenamefont
  {Yi}}]{ManyBodyNHSEgaugecoupling2023}%
  \BibitemOpen
  \bibfield  {author} {\bibinfo {author} {\bibfnamefont {H.}~\bibnamefont
  {Li}}, \bibinfo {author} {\bibfnamefont {H.}~\bibnamefont {Wu}}, \bibinfo
  {author} {\bibfnamefont {W.}~\bibnamefont {Zheng}},\ and\ \bibinfo {author}
  {\bibfnamefont {W.}~\bibnamefont {Yi}},\ }\bibfield  {title} {\bibinfo
  {title} {Many-body non-hermitian skin effect under dynamic gauge coupling},\
  }\href {https://doi.org/10.1103/PhysRevResearch.5.033173} {\bibfield
  {journal} {\bibinfo  {journal} {Phys. Rev. Res.}\ }\textbf {\bibinfo {volume}
  {5}},\ \bibinfo {pages} {033173} (\bibinfo {year}
  {2023}{\natexlab{a}})}\BibitemShut {NoStop}%
\bibitem [{\citenamefont {Haga}\ \emph {et~al.}(2021)\citenamefont {Haga},
  \citenamefont {Nakagawa}, \citenamefont {Hamazaki},\ and\ \citenamefont
  {Ueda}}]{hagaLiouvillianSkinEffect2021}%
  \BibitemOpen
  \bibfield  {author} {\bibinfo {author} {\bibfnamefont {T.}~\bibnamefont
  {Haga}}, \bibinfo {author} {\bibfnamefont {M.}~\bibnamefont {Nakagawa}},
  \bibinfo {author} {\bibfnamefont {R.}~\bibnamefont {Hamazaki}},\ and\
  \bibinfo {author} {\bibfnamefont {M.}~\bibnamefont {Ueda}},\ }\bibfield
  {title} {\bibinfo {title} {Liouvillian skin effect: Slowing down of
  relaxation processes without gap closing},\ }\href
  {https://doi.org/10.1103/PhysRevLett.127.070402} {\bibfield  {journal}
  {\bibinfo  {journal} {Phys. Rev. Lett.}\ }\textbf {\bibinfo {volume} {127}},\
  \bibinfo {pages} {070402} (\bibinfo {year} {2021})}\BibitemShut {NoStop}%
\bibitem [{\citenamefont {Mao}\ \emph {et~al.}(2024)\citenamefont {Mao},
  \citenamefont {Tao}, \citenamefont {Hu},\ and\ \citenamefont
  {Pan}}]{mao-liouvillian2024}%
  \BibitemOpen
  \bibfield  {author} {\bibinfo {author} {\bibfnamefont {L.}~\bibnamefont
  {Mao}}, \bibinfo {author} {\bibfnamefont {M.-J.}\ \bibnamefont {Tao}},
  \bibinfo {author} {\bibfnamefont {H.}~\bibnamefont {Hu}},\ and\ \bibinfo
  {author} {\bibfnamefont {L.}~\bibnamefont {Pan}},\ }\bibfield  {title}
  {\bibinfo {title} {Liouvillian skin effect in a one-dimensional open
  many-body quantum system with generalized boundary conditions},\ }\href
  {http://arxiv.org/abs/2401.15614} {\bibfield  {journal} {\bibinfo  {journal}
  {arXiv:2401.15614}\ } (\bibinfo {year} {2024})}\BibitemShut {NoStop}%
\bibitem [{\citenamefont {Yang}\ \emph {et~al.}(2022)\citenamefont {Yang},
  \citenamefont {Jiang},\ and\ \citenamefont {Bergholtz}}]{Liouvillianskin}%
  \BibitemOpen
  \bibfield  {author} {\bibinfo {author} {\bibfnamefont {F.}~\bibnamefont
  {Yang}}, \bibinfo {author} {\bibfnamefont {Q.-D.}\ \bibnamefont {Jiang}},\
  and\ \bibinfo {author} {\bibfnamefont {E.~J.}\ \bibnamefont {Bergholtz}},\
  }\bibfield  {title} {\bibinfo {title} {Liouvillian skin effect in an exactly
  solvable model},\ }\href {https://doi.org/10.1103/PhysRevResearch.4.023160}
  {\bibfield  {journal} {\bibinfo  {journal} {Phys. Rev. Res.}\ }\textbf
  {\bibinfo {volume} {4}},\ \bibinfo {pages} {023160} (\bibinfo {year}
  {2022})}\BibitemShut {NoStop}%
\bibitem [{\citenamefont {Song}\ \emph {et~al.}(2019)\citenamefont {Song},
  \citenamefont {Yao},\ and\ \citenamefont {Wang}}]{chiraldamping}%
  \BibitemOpen
  \bibfield  {author} {\bibinfo {author} {\bibfnamefont {F.}~\bibnamefont
  {Song}}, \bibinfo {author} {\bibfnamefont {S.}~\bibnamefont {Yao}},\ and\
  \bibinfo {author} {\bibfnamefont {Z.}~\bibnamefont {Wang}},\ }\bibfield
  {title} {\bibinfo {title} {Non-hermitian skin effect and chiral damping in
  open quantum systems},\ }\href
  {https://doi.org/10.1103/PhysRevLett.123.170401} {\bibfield  {journal}
  {\bibinfo  {journal} {Phys. Rev. Lett.}\ }\textbf {\bibinfo {volume} {123}},\
  \bibinfo {pages} {170401} (\bibinfo {year} {2019})}\BibitemShut {NoStop}%
\bibitem [{\citenamefont {Liu}\ \emph {et~al.}(2020)\citenamefont {Liu},
  \citenamefont {Zhang}, \citenamefont {Yang},\ and\ \citenamefont
  {Chen}}]{helicaldamping}%
  \BibitemOpen
  \bibfield  {author} {\bibinfo {author} {\bibfnamefont {C.-H.}\ \bibnamefont
  {Liu}}, \bibinfo {author} {\bibfnamefont {K.}~\bibnamefont {Zhang}}, \bibinfo
  {author} {\bibfnamefont {Z.}~\bibnamefont {Yang}},\ and\ \bibinfo {author}
  {\bibfnamefont {S.}~\bibnamefont {Chen}},\ }\bibfield  {title} {\bibinfo
  {title} {Helical damping and dynamical critical skin effect in open quantum
  systems},\ }\href {https://doi.org/10.1103/PhysRevResearch.2.043167}
  {\bibfield  {journal} {\bibinfo  {journal} {Phys. Rev. Research}\ }\textbf
  {\bibinfo {volume} {2}},\ \bibinfo {pages} {043167} (\bibinfo {year}
  {2020})}\BibitemShut {NoStop}%
\bibitem [{\citenamefont {Wang}\ \emph {et~al.}(2023)\citenamefont {Wang},
  \citenamefont {Lu}, \citenamefont {Peng}, \citenamefont {Qi}, \citenamefont
  {Wang},\ and\ \citenamefont {Jie}}]{wangAcceleratingRelaxationDynamics2023}%
  \BibitemOpen
  \bibfield  {author} {\bibinfo {author} {\bibfnamefont {Z.}~\bibnamefont
  {Wang}}, \bibinfo {author} {\bibfnamefont {Y.}~\bibnamefont {Lu}}, \bibinfo
  {author} {\bibfnamefont {Y.}~\bibnamefont {Peng}}, \bibinfo {author}
  {\bibfnamefont {R.}~\bibnamefont {Qi}}, \bibinfo {author} {\bibfnamefont
  {Y.}~\bibnamefont {Wang}},\ and\ \bibinfo {author} {\bibfnamefont
  {J.}~\bibnamefont {Jie}},\ }\bibfield  {title} {\bibinfo {title}
  {Accelerating relaxation dynamics in open quantum systems with liouvillian
  skin effect},\ }\href {https://doi.org/10.1103/PhysRevB.108.054313}
  {\bibfield  {journal} {\bibinfo  {journal} {Phys. Rev. B}\ }\textbf {\bibinfo
  {volume} {108}},\ \bibinfo {pages} {054313} (\bibinfo {year}
  {2023})}\BibitemShut {NoStop}%
\bibitem [{\citenamefont {Mori}\ and\ \citenamefont
  {Shirai}(2023)}]{symmetrizedLiouvillianGap}%
  \BibitemOpen
  \bibfield  {author} {\bibinfo {author} {\bibfnamefont {T.}~\bibnamefont
  {Mori}}\ and\ \bibinfo {author} {\bibfnamefont {T.}~\bibnamefont {Shirai}},\
  }\bibfield  {title} {\bibinfo {title} {Symmetrized liouvillian gap in
  markovian open quantum systems},\ }\href
  {https://doi.org/10.1103/PhysRevLett.130.230404} {\bibfield  {journal}
  {\bibinfo  {journal} {Phys. Rev. Lett.}\ }\textbf {\bibinfo {volume} {130}},\
  \bibinfo {pages} {230404} (\bibinfo {year} {2023})}\BibitemShut {NoStop}%
\bibitem [{\citenamefont {\ifmmode \check{Z}\else
  \v{Z}\fi{}nidari\ifmmode~\check{c}\else
  \v{c}\fi{}}(2023)}]{Phantomrelaxation}%
  \BibitemOpen
  \bibfield  {author} {\bibinfo {author} {\bibfnamefont {M.}~\bibnamefont
  {\ifmmode \check{Z}\else \v{Z}\fi{}nidari\ifmmode~\check{c}\else
  \v{c}\fi{}}},\ }\bibfield  {title} {\bibinfo {title} {Phantom relaxation rate
  of the average purity evolution in random circuits due to jordan
  non-hermitian skin effect and magic sums},\ }\href
  {https://doi.org/10.1103/PhysRevResearch.5.033145} {\bibfield  {journal}
  {\bibinfo  {journal} {Phys. Rev. Res.}\ }\textbf {\bibinfo {volume} {5}},\
  \bibinfo {pages} {033145} (\bibinfo {year} {2023})}\BibitemShut {NoStop}%
\bibitem [{\citenamefont {Begg}\ and\ \citenamefont
  {Hanai}(2024)}]{QuantumCriticalitynonreciprocity}%
  \BibitemOpen
  \bibfield  {author} {\bibinfo {author} {\bibfnamefont {S.~E.}\ \bibnamefont
  {Begg}}\ and\ \bibinfo {author} {\bibfnamefont {R.}~\bibnamefont {Hanai}},\
  }\bibfield  {title} {\bibinfo {title} {Quantum criticality in open quantum
  spin chains with nonreciprocity},\ }\href
  {https://doi.org/10.1103/PhysRevLett.132.120401} {\bibfield  {journal}
  {\bibinfo  {journal} {Phys. Rev. Lett.}\ }\textbf {\bibinfo {volume} {132}},\
  \bibinfo {pages} {120401} (\bibinfo {year} {2024})}\BibitemShut {NoStop}%
\bibitem [{\citenamefont {Garbe}\ \emph {et~al.}(2024)\citenamefont {Garbe},
  \citenamefont {Minoguchi}, \citenamefont {Huber},\ and\ \citenamefont
  {Rabl}}]{BosonicLSE}%
  \BibitemOpen
  \bibfield  {author} {\bibinfo {author} {\bibfnamefont {L.}~\bibnamefont
  {Garbe}}, \bibinfo {author} {\bibfnamefont {Y.}~\bibnamefont {Minoguchi}},
  \bibinfo {author} {\bibfnamefont {J.}~\bibnamefont {Huber}},\ and\ \bibinfo
  {author} {\bibfnamefont {P.}~\bibnamefont {Rabl}},\ }\bibfield  {title}
  {\bibinfo {title} {{The bosonic skin effect: Boundary condensation in
  asymmetric transport}},\ }\href
  {https://doi.org/10.21468/SciPostPhys.16.1.029} {\bibfield  {journal}
  {\bibinfo  {journal} {SciPost Phys.}\ }\textbf {\bibinfo {volume} {16}},\
  \bibinfo {pages} {029} (\bibinfo {year} {2024})}\BibitemShut {NoStop}%
\bibitem [{\citenamefont {Wang}\ \emph
  {et~al.}(2024{\natexlab{b}})\citenamefont {Wang}, \citenamefont {Fang},\ and\
  \citenamefont {Ren}}]{MISE}%
  \BibitemOpen
  \bibfield  {author} {\bibinfo {author} {\bibfnamefont {Y.-P.}\ \bibnamefont
  {Wang}}, \bibinfo {author} {\bibfnamefont {C.}~\bibnamefont {Fang}},\ and\
  \bibinfo {author} {\bibfnamefont {J.}~\bibnamefont {Ren}},\ }\bibfield
  {title} {\bibinfo {title} {Absence of measurement-induced entanglement
  transition due to feedback-induced skin effect},\ }\href
  {https://doi.org/10.1103/PhysRevB.110.035113} {\bibfield  {journal} {\bibinfo
   {journal} {Phys. Rev. B}\ }\textbf {\bibinfo {volume} {110}},\ \bibinfo
  {pages} {035113} (\bibinfo {year} {2024}{\natexlab{b}})}\BibitemShut
  {NoStop}%
\bibitem [{\citenamefont {Feng}\ \emph {et~al.}(2023)\citenamefont {Feng},
  \citenamefont {Liu}, \citenamefont {Chen},\ and\ \citenamefont
  {Guo}}]{MISE-longrange}%
  \BibitemOpen
  \bibfield  {author} {\bibinfo {author} {\bibfnamefont {X.}~\bibnamefont
  {Feng}}, \bibinfo {author} {\bibfnamefont {S.}~\bibnamefont {Liu}}, \bibinfo
  {author} {\bibfnamefont {S.}~\bibnamefont {Chen}},\ and\ \bibinfo {author}
  {\bibfnamefont {W.}~\bibnamefont {Guo}},\ }\bibfield  {title} {\bibinfo
  {title} {Absence of logarithmic and algebraic scaling entanglement phases due
  to the skin effect},\ }\href {https://doi.org/10.1103/PhysRevB.107.094309}
  {\bibfield  {journal} {\bibinfo  {journal} {Phys. Rev. B}\ }\textbf {\bibinfo
  {volume} {107}},\ \bibinfo {pages} {094309} (\bibinfo {year}
  {2023})}\BibitemShut {NoStop}%
\bibitem [{\citenamefont {Skinner}\ \emph {et~al.}(2019)\citenamefont
  {Skinner}, \citenamefont {Ruhman},\ and\ \citenamefont {Nahum}}]{Nahum}%
  \BibitemOpen
  \bibfield  {author} {\bibinfo {author} {\bibfnamefont {B.}~\bibnamefont
  {Skinner}}, \bibinfo {author} {\bibfnamefont {J.}~\bibnamefont {Ruhman}},\
  and\ \bibinfo {author} {\bibfnamefont {A.}~\bibnamefont {Nahum}},\ }\bibfield
   {title} {\bibinfo {title} {Measurement-induced phase transitions in the
  dynamics of entanglement},\ }\href
  {https://doi.org/10.1103/PhysRevX.9.031009} {\bibfield  {journal} {\bibinfo
  {journal} {Phys. Rev. X}\ }\textbf {\bibinfo {volume} {9}},\ \bibinfo {pages}
  {031009} (\bibinfo {year} {2019})}\BibitemShut {NoStop}%
\bibitem [{\citenamefont {Li}\ \emph {et~al.}(2018)\citenamefont {Li},
  \citenamefont {Chen},\ and\ \citenamefont {Fisher}}]{FisherPRB1}%
  \BibitemOpen
  \bibfield  {author} {\bibinfo {author} {\bibfnamefont {Y.}~\bibnamefont
  {Li}}, \bibinfo {author} {\bibfnamefont {X.}~\bibnamefont {Chen}},\ and\
  \bibinfo {author} {\bibfnamefont {M.~P.~A.}\ \bibnamefont {Fisher}},\
  }\bibfield  {title} {\bibinfo {title} {Quantum zeno effect and the many-body
  entanglement transition},\ }\href
  {https://doi.org/10.1103/PhysRevB.98.205136} {\bibfield  {journal} {\bibinfo
  {journal} {Phys. Rev. B}\ }\textbf {\bibinfo {volume} {98}},\ \bibinfo
  {pages} {205136} (\bibinfo {year} {2018})}\BibitemShut {NoStop}%
\bibitem [{\citenamefont {Li}\ \emph {et~al.}(2019)\citenamefont {Li},
  \citenamefont {Chen},\ and\ \citenamefont {Fisher}}]{FishePRB2}%
  \BibitemOpen
  \bibfield  {author} {\bibinfo {author} {\bibfnamefont {Y.}~\bibnamefont
  {Li}}, \bibinfo {author} {\bibfnamefont {X.}~\bibnamefont {Chen}},\ and\
  \bibinfo {author} {\bibfnamefont {M.~P.~A.}\ \bibnamefont {Fisher}},\
  }\bibfield  {title} {\bibinfo {title} {Measurement-driven entanglement
  transition in hybrid quantum circuits},\ }\href
  {https://doi.org/10.1103/PhysRevB.100.134306} {\bibfield  {journal} {\bibinfo
   {journal} {Phys. Rev. B}\ }\textbf {\bibinfo {volume} {100}},\ \bibinfo
  {pages} {134306} (\bibinfo {year} {2019})}\BibitemShut {NoStop}%
\bibitem [{\citenamefont {Cao}\ \emph {et~al.}(2019)\citenamefont {Cao},
  \citenamefont {Tilloy},\ and\ \citenamefont {Luca}}]{fermionMIPT}%
  \BibitemOpen
  \bibfield  {author} {\bibinfo {author} {\bibfnamefont {X.}~\bibnamefont
  {Cao}}, \bibinfo {author} {\bibfnamefont {A.}~\bibnamefont {Tilloy}},\ and\
  \bibinfo {author} {\bibfnamefont {A.~D.}\ \bibnamefont {Luca}},\ }\bibfield
  {title} {\bibinfo {title} {{Entanglement in a fermion chain under continuous
  monitoring}},\ }\href {https://doi.org/10.21468/SciPostPhys.7.2.024}
  {\bibfield  {journal} {\bibinfo  {journal} {SciPost Phys.}\ }\textbf
  {\bibinfo {volume} {7}},\ \bibinfo {pages} {024} (\bibinfo {year}
  {2019})}\BibitemShut {NoStop}%
\bibitem [{\citenamefont {Alberton}\ \emph {et~al.}(2021)\citenamefont
  {Alberton}, \citenamefont {Buchhold},\ and\ \citenamefont
  {Diehl}}]{DiehlMIPT}%
  \BibitemOpen
  \bibfield  {author} {\bibinfo {author} {\bibfnamefont {O.}~\bibnamefont
  {Alberton}}, \bibinfo {author} {\bibfnamefont {M.}~\bibnamefont {Buchhold}},\
  and\ \bibinfo {author} {\bibfnamefont {S.}~\bibnamefont {Diehl}},\ }\bibfield
   {title} {\bibinfo {title} {Entanglement transition in a monitored
  free-fermion chain: From extended criticality to area law},\ }\href
  {https://doi.org/10.1103/PhysRevLett.126.170602} {\bibfield  {journal}
  {\bibinfo  {journal} {Phys. Rev. Lett.}\ }\textbf {\bibinfo {volume} {126}},\
  \bibinfo {pages} {170602} (\bibinfo {year} {2021})}\BibitemShut {NoStop}%
\bibitem [{\citenamefont {M\"uller}\ \emph {et~al.}(2022)\citenamefont
  {M\"uller}, \citenamefont {Diehl},\ and\ \citenamefont
  {Buchhold}}]{longrangeMIPT}%
  \BibitemOpen
  \bibfield  {author} {\bibinfo {author} {\bibfnamefont {T.}~\bibnamefont
  {M\"uller}}, \bibinfo {author} {\bibfnamefont {S.}~\bibnamefont {Diehl}},\
  and\ \bibinfo {author} {\bibfnamefont {M.}~\bibnamefont {Buchhold}},\
  }\bibfield  {title} {\bibinfo {title} {Measurement-induced dark state phase
  transitions in long-ranged fermion systems},\ }\href
  {https://doi.org/10.1103/PhysRevLett.128.010605} {\bibfield  {journal}
  {\bibinfo  {journal} {Phys. Rev. Lett.}\ }\textbf {\bibinfo {volume} {128}},\
  \bibinfo {pages} {010605} (\bibinfo {year} {2022})}\BibitemShut {NoStop}%
\bibitem [{\citenamefont {Poboiko}\ \emph {et~al.}(2024)\citenamefont
  {Poboiko}, \citenamefont {Gornyi},\ and\ \citenamefont
  {Mirlin}}]{MIPTtwodimension}%
  \BibitemOpen
  \bibfield  {author} {\bibinfo {author} {\bibfnamefont {I.}~\bibnamefont
  {Poboiko}}, \bibinfo {author} {\bibfnamefont {I.~V.}\ \bibnamefont
  {Gornyi}},\ and\ \bibinfo {author} {\bibfnamefont {A.~D.}\ \bibnamefont
  {Mirlin}},\ }\bibfield  {title} {\bibinfo {title} {Measurement-induced phase
  transition for free fermions above one dimension},\ }\href
  {https://doi.org/10.1103/PhysRevLett.132.110403} {\bibfield  {journal}
  {\bibinfo  {journal} {Phys. Rev. Lett.}\ }\textbf {\bibinfo {volume} {132}},\
  \bibinfo {pages} {110403} (\bibinfo {year} {2024})}\BibitemShut {NoStop}%
\bibitem [{\citenamefont {Liu}\ \emph {et~al.}(2023)\citenamefont {Liu},
  \citenamefont {Li}, \citenamefont {Zhang}, \citenamefont {Jian},\ and\
  \citenamefont {Yao}}]{MIPTLiu1}%
  \BibitemOpen
  \bibfield  {author} {\bibinfo {author} {\bibfnamefont {S.}~\bibnamefont
  {Liu}}, \bibinfo {author} {\bibfnamefont {M.-R.}\ \bibnamefont {Li}},
  \bibinfo {author} {\bibfnamefont {S.-X.}\ \bibnamefont {Zhang}}, \bibinfo
  {author} {\bibfnamefont {S.-K.}\ \bibnamefont {Jian}},\ and\ \bibinfo
  {author} {\bibfnamefont {H.}~\bibnamefont {Yao}},\ }\bibfield  {title}
  {\bibinfo {title} {Universal kardar-parisi-zhang scaling in noisy hybrid
  quantum circuits},\ }\href {https://doi.org/10.1103/PhysRevB.107.L201113}
  {\bibfield  {journal} {\bibinfo  {journal} {Phys. Rev. B}\ }\textbf {\bibinfo
  {volume} {107}},\ \bibinfo {pages} {L201113} (\bibinfo {year}
  {2023})}\BibitemShut {NoStop}%
\bibitem [{\citenamefont {Liu}\ \emph {et~al.}(2024{\natexlab{a}})\citenamefont
  {Liu}, \citenamefont {Li}, \citenamefont {Zhang},\ and\ \citenamefont
  {Jian}}]{MIPTLiu2}%
  \BibitemOpen
  \bibfield  {author} {\bibinfo {author} {\bibfnamefont {S.}~\bibnamefont
  {Liu}}, \bibinfo {author} {\bibfnamefont {M.-R.}\ \bibnamefont {Li}},
  \bibinfo {author} {\bibfnamefont {S.-X.}\ \bibnamefont {Zhang}},\ and\
  \bibinfo {author} {\bibfnamefont {S.-K.}\ \bibnamefont {Jian}},\ }\bibfield
  {title} {\bibinfo {title} {Entanglement structure and information protection
  in noisy hybrid quantum circuits},\ }\href
  {https://doi.org/10.1103/PhysRevLett.132.240402} {\bibfield  {journal}
  {\bibinfo  {journal} {Phys. Rev. Lett.}\ }\textbf {\bibinfo {volume} {132}},\
  \bibinfo {pages} {240402} (\bibinfo {year} {2024}{\natexlab{a}})}\BibitemShut
  {NoStop}%
\bibitem [{\citenamefont {Liu}\ \emph {et~al.}(2024{\natexlab{b}})\citenamefont
  {Liu}, \citenamefont {Li}, \citenamefont {Zhang}, \citenamefont {Jian},\ and\
  \citenamefont {Yao}}]{MIPTLiu3}%
  \BibitemOpen
  \bibfield  {author} {\bibinfo {author} {\bibfnamefont {S.}~\bibnamefont
  {Liu}}, \bibinfo {author} {\bibfnamefont {M.-R.}\ \bibnamefont {Li}},
  \bibinfo {author} {\bibfnamefont {S.-X.}\ \bibnamefont {Zhang}}, \bibinfo
  {author} {\bibfnamefont {S.-K.}\ \bibnamefont {Jian}},\ and\ \bibinfo
  {author} {\bibfnamefont {H.}~\bibnamefont {Yao}},\ }\bibfield  {title}
  {\bibinfo {title} {Noise-induced phase transitions in hybrid quantum
  circuits},\ }\href {https://arxiv.org/pdf/2401.16631} {\bibfield  {journal}
  {\bibinfo  {journal} {arXiv:2401.16631}\ } (\bibinfo {year}
  {2024}{\natexlab{b}})}\BibitemShut {NoStop}%
\bibitem [{\citenamefont {Gullans}\ and\ \citenamefont
  {Huse}(2019)}]{EPTAnderson}%
  \BibitemOpen
  \bibfield  {author} {\bibinfo {author} {\bibfnamefont {M.~J.}\ \bibnamefont
  {Gullans}}\ and\ \bibinfo {author} {\bibfnamefont {D.~A.}\ \bibnamefont
  {Huse}},\ }\bibfield  {title} {\bibinfo {title} {Localization as an
  entanglement phase transition in boundary-driven anderson models},\ }\href
  {https://doi.org/10.1103/PhysRevLett.123.110601} {\bibfield  {journal}
  {\bibinfo  {journal} {Phys. Rev. Lett.}\ }\textbf {\bibinfo {volume} {123}},\
  \bibinfo {pages} {110601} (\bibinfo {year} {2019})}\BibitemShut {NoStop}%
\bibitem [{\citenamefont {de~Albornoz}\ \emph {et~al.}(2024)\citenamefont
  {de~Albornoz}, \citenamefont {Rose},\ and\ \citenamefont
  {Pal}}]{EPTLocalizationlongRange}%
  \BibitemOpen
  \bibfield  {author} {\bibinfo {author} {\bibfnamefont {A.~C.~C.}\
  \bibnamefont {de~Albornoz}}, \bibinfo {author} {\bibfnamefont {D.~C.}\
  \bibnamefont {Rose}},\ and\ \bibinfo {author} {\bibfnamefont
  {A.}~\bibnamefont {Pal}},\ }\bibfield  {title} {\bibinfo {title}
  {Entanglement transition and heterogeneity in long-range quadratic
  lindbladians},\ }\href {https://doi.org/10.1103/PhysRevB.109.214204}
  {\bibfield  {journal} {\bibinfo  {journal} {Phys. Rev. B}\ }\textbf {\bibinfo
  {volume} {109}},\ \bibinfo {pages} {214204} (\bibinfo {year}
  {2024})}\BibitemShut {NoStop}%
\bibitem [{\citenamefont {Kawabata}\ \emph {et~al.}(2023)\citenamefont
  {Kawabata}, \citenamefont {Numasawa},\ and\ \citenamefont
  {Ryu}}]{EPTNonHermitian}%
  \BibitemOpen
  \bibfield  {author} {\bibinfo {author} {\bibfnamefont {K.}~\bibnamefont
  {Kawabata}}, \bibinfo {author} {\bibfnamefont {T.}~\bibnamefont {Numasawa}},\
  and\ \bibinfo {author} {\bibfnamefont {S.}~\bibnamefont {Ryu}},\ }\bibfield
  {title} {\bibinfo {title} {Entanglement phase transition induced by the
  non-hermitian skin effect},\ }\href
  {https://doi.org/10.1103/PhysRevX.13.021007} {\bibfield  {journal} {\bibinfo
  {journal} {Phys. Rev. X}\ }\textbf {\bibinfo {volume} {13}},\ \bibinfo
  {pages} {021007} (\bibinfo {year} {2023})}\BibitemShut {NoStop}%
\bibitem [{\citenamefont {Li}\ \emph {et~al.}(2023{\natexlab{b}})\citenamefont
  {Li}, \citenamefont {Liu},\ and\ \citenamefont
  {Xu}}]{liDisorderInducedEntanglementPhase2023}%
  \BibitemOpen
  \bibfield  {author} {\bibinfo {author} {\bibfnamefont {K.}~\bibnamefont
  {Li}}, \bibinfo {author} {\bibfnamefont {Z.-C.}\ \bibnamefont {Liu}},\ and\
  \bibinfo {author} {\bibfnamefont {Y.}~\bibnamefont {Xu}},\ }\bibfield
  {title} {\bibinfo {title} {Disorder-induced entanglement phase transitions in
  non-hermitian systems with skin effects},\ }\href
  {https://arxiv.org/abs/2305.12342} {\bibfield  {journal} {\bibinfo  {journal}
  {arXiv:2305.12342}\ } (\bibinfo {year} {2023}{\natexlab{b}})}\BibitemShut
  {NoStop}%
\bibitem [{\citenamefont {Zhou}(2024)}]{NHEvoQuasiDisorder}%
  \BibitemOpen
  \bibfield  {author} {\bibinfo {author} {\bibfnamefont {L.}~\bibnamefont
  {Zhou}},\ }\bibfield  {title} {\bibinfo {title} {Entanglement phase
  transitions in non-hermitian quasicrystals},\ }\href
  {https://doi.org/10.1103/PhysRevB.109.024204} {\bibfield  {journal} {\bibinfo
   {journal} {Phys. Rev. B}\ }\textbf {\bibinfo {volume} {109}},\ \bibinfo
  {pages} {024204} (\bibinfo {year} {2024})}\BibitemShut {NoStop}%
\bibitem [{\citenamefont {Li}\ \emph {et~al.}(2024)\citenamefont {Li},
  \citenamefont {Yu},\ and\ \citenamefont {Li}}]{NHEvoAA}%
  \BibitemOpen
  \bibfield  {author} {\bibinfo {author} {\bibfnamefont {S.-Z.}\ \bibnamefont
  {Li}}, \bibinfo {author} {\bibfnamefont {X.-J.}\ \bibnamefont {Yu}},\ and\
  \bibinfo {author} {\bibfnamefont {Z.}~\bibnamefont {Li}},\ }\bibfield
  {title} {\bibinfo {title} {Emergent entanglement phase transitions in
  non-hermitian aubry-andr\'e-harper chains},\ }\href
  {https://doi.org/10.1103/PhysRevB.109.024306} {\bibfield  {journal} {\bibinfo
   {journal} {Phys. Rev. B}\ }\textbf {\bibinfo {volume} {109}},\ \bibinfo
  {pages} {024306} (\bibinfo {year} {2024})}\BibitemShut {NoStop}%
\bibitem [{\citenamefont {Liu}\ and\ \citenamefont
  {Chen}(2024)}]{liu-nonlinear-2024}%
  \BibitemOpen
  \bibfield  {author} {\bibinfo {author} {\bibfnamefont {Y.-G.}\ \bibnamefont
  {Liu}}\ and\ \bibinfo {author} {\bibfnamefont {S.}~\bibnamefont {Chen}},\
  }\bibfield  {title} {\bibinfo {title} {Nonlinear lindblad master equation and
  postselected skin effect},\ }\href {http://arxiv.org/abs/2405.11812}
  {\bibfield  {journal} {\bibinfo  {journal} {arXiv:2405.11812}\ } (\bibinfo
  {year} {2024})}\BibitemShut {NoStop}%
\bibitem [{\citenamefont {Liu}\ \emph {et~al.}(2024{\natexlab{c}})\citenamefont
  {Liu}, \citenamefont {Li},\ and\ \citenamefont {Xu}}]{MISE-DPT}%
  \BibitemOpen
  \bibfield  {author} {\bibinfo {author} {\bibfnamefont {Z.-C.}\ \bibnamefont
  {Liu}}, \bibinfo {author} {\bibfnamefont {K.}~\bibnamefont {Li}},\ and\
  \bibinfo {author} {\bibfnamefont {Y.}~\bibnamefont {Xu}},\ }\bibfield
  {title} {\bibinfo {title} {Dynamical {{Phase Transition}} due to
  {{Feedback-induced Skin Effect}}},\ }\href {https://arxiv.org/abs/2311.16541}
  {\bibfield  {journal} {\bibinfo  {journal} {arXiv:2311.16541}\ } (\bibinfo
  {year} {2024}{\natexlab{c}})}\BibitemShut {NoStop}%
\bibitem [{\citenamefont {Hatano}\ and\ \citenamefont
  {Nelson}(1996)}]{HNmodel}%
  \BibitemOpen
  \bibfield  {author} {\bibinfo {author} {\bibfnamefont {N.}~\bibnamefont
  {Hatano}}\ and\ \bibinfo {author} {\bibfnamefont {D.~R.}\ \bibnamefont
  {Nelson}},\ }\bibfield  {title} {\bibinfo {title} {Localization transitions
  in non-hermitian quantum mechanics},\ }\href
  {https://doi.org/10.1103/PhysRevLett.77.570} {\bibfield  {journal} {\bibinfo
  {journal} {Phys. Rev. Lett.}\ }\textbf {\bibinfo {volume} {77}},\ \bibinfo
  {pages} {570} (\bibinfo {year} {1996})}\BibitemShut {NoStop}%
\bibitem [{\citenamefont {Li}\ \emph {et~al.}(2020)\citenamefont {Li},
  \citenamefont {Lee}, \citenamefont {Mu},\ and\ \citenamefont {Gong}}]{CNHSE}%
  \BibitemOpen
  \bibfield  {author} {\bibinfo {author} {\bibfnamefont {L.}~\bibnamefont
  {Li}}, \bibinfo {author} {\bibfnamefont {C.~H.}\ \bibnamefont {Lee}},
  \bibinfo {author} {\bibfnamefont {S.}~\bibnamefont {Mu}},\ and\ \bibinfo
  {author} {\bibfnamefont {J.}~\bibnamefont {Gong}},\ }\bibfield  {title}
  {\bibinfo {title} {Critical non-hermitian skin effect},\ }\href
  {https://doi.org/10.1038/s41467-020-18917-4} {\bibfield  {journal} {\bibinfo
  {journal} {Nature communications}\ }\textbf {\bibinfo {volume} {11}},\
  \bibinfo {pages} {1} (\bibinfo {year} {2020})}\BibitemShut {NoStop}%
\bibitem [{\citenamefont {Yokomizo}\ and\ \citenamefont
  {Murakami}(2021)}]{scalingCNHSE}%
  \BibitemOpen
  \bibfield  {author} {\bibinfo {author} {\bibfnamefont {K.}~\bibnamefont
  {Yokomizo}}\ and\ \bibinfo {author} {\bibfnamefont {S.}~\bibnamefont
  {Murakami}},\ }\bibfield  {title} {\bibinfo {title} {Scaling rule for the
  critical non-hermitian skin effect},\ }\href
  {https://doi.org/10.1103/PhysRevB.104.165117} {\bibfield  {journal} {\bibinfo
   {journal} {Phys. Rev. B}\ }\textbf {\bibinfo {volume} {104}},\ \bibinfo
  {pages} {165117} (\bibinfo {year} {2021})}\BibitemShut {NoStop}%
\bibitem [{\citenamefont {Wiseman}\ and\ \citenamefont
  {Milburn}(2009)}]{quantum-measurement-and-control}%
  \BibitemOpen
  \bibfield  {author} {\bibinfo {author} {\bibfnamefont {H.~M.}\ \bibnamefont
  {Wiseman}}\ and\ \bibinfo {author} {\bibfnamefont {G.~J.}\ \bibnamefont
  {Milburn}},\ }\bibfield  {title} {\bibinfo {title} {Quantum measurement and
  control}\ }\href {https://doi.org/10.1017/CBO9780511813948}
  {10.1017/CBO9780511813948} (\bibinfo {year} {2009})\BibitemShut {NoStop}%
\bibitem [{\citenamefont {Jacobs}\ and\ \citenamefont
  {Steck}(2006)}]{continuous-quantum-measurement}%
  \BibitemOpen
  \bibfield  {author} {\bibinfo {author} {\bibfnamefont {K.}~\bibnamefont
  {Jacobs}}\ and\ \bibinfo {author} {\bibfnamefont {D.~A.}\ \bibnamefont
  {Steck}},\ }\bibfield  {title} {\bibinfo {title} {A straightforward
  introduction to continuous quantum measurement},\ }\href
  {https://doi.org/10.1080/00107510601101934} {\bibfield  {journal} {\bibinfo
  {journal} {Contemporary Physics}\ }\textbf {\bibinfo {volume} {47}},\
  \bibinfo {pages} {279} (\bibinfo {year} {2006})}\BibitemShut {NoStop}%
\bibitem [{\citenamefont {Gardiner}\ and\ \citenamefont
  {Zoller}(2004)}]{quantum-noise}%
  \BibitemOpen
  \bibfield  {author} {\bibinfo {author} {\bibfnamefont {C.}~\bibnamefont
  {Gardiner}}\ and\ \bibinfo {author} {\bibfnamefont {P.}~\bibnamefont
  {Zoller}},\ }\bibfield  {title} {\bibinfo {title} {Quantum noise: A handbook
  of markovian and non-markovian quantum stochastic methods with applications
  to quantum optics},\ }\href {https://link.springer.com/book/9783540223016} {\
  \bibinfo {series} {Springer Series in Synergetics} (\bibinfo {year}
  {2004})}\BibitemShut {NoStop}%
\bibitem [{\citenamefont {Fruchart}\ \emph {et~al.}(2021)\citenamefont
  {Fruchart}, \citenamefont {Hanai}, \citenamefont {Littlewood},\ and\
  \citenamefont {Vitelli}}]{non-reciprocal-phasetransition}%
  \BibitemOpen
  \bibfield  {author} {\bibinfo {author} {\bibfnamefont {M.}~\bibnamefont
  {Fruchart}}, \bibinfo {author} {\bibfnamefont {R.}~\bibnamefont {Hanai}},
  \bibinfo {author} {\bibfnamefont {P.~B.}\ \bibnamefont {Littlewood}},\ and\
  \bibinfo {author} {\bibfnamefont {V.}~\bibnamefont {Vitelli}},\ }\bibfield
  {title} {\bibinfo {title} {Non-reciprocal phase transitions},\ }\href
  {https://doi.org/10.1038/s41586-021-03375-9} {\bibfield  {journal} {\bibinfo
  {journal} {Nature}\ }\textbf {\bibinfo {volume} {592}},\ \bibinfo {pages}
  {363} (\bibinfo {year} {2021})}\BibitemShut {NoStop}%
\bibitem [{\citenamefont {Choi}(1975)}]{CHOI}%
  \BibitemOpen
  \bibfield  {author} {\bibinfo {author} {\bibfnamefont {M.-D.}\ \bibnamefont
  {Choi}},\ }\bibfield  {title} {\bibinfo {title} {Completely positive linear
  maps on complex matrices},\ }\href
  {https://doi.org/https://doi.org/10.1016/0024-3795(75)90075-0} {\bibfield
  {journal} {\bibinfo  {journal} {Linear Algebra and its Applications}\
  }\textbf {\bibinfo {volume} {10}},\ \bibinfo {pages} {285} (\bibinfo {year}
  {1975})}\BibitemShut {NoStop}%
\bibitem [{\citenamefont {Jamiołkowski}(1972)}]{JAMIOLKOWSKI}%
  \BibitemOpen
  \bibfield  {author} {\bibinfo {author} {\bibfnamefont {A.}~\bibnamefont
  {Jamiołkowski}},\ }\bibfield  {title} {\bibinfo {title} {Linear
  transformations which preserve trace and positive semidefiniteness of
  operators},\ }\href
  {https://doi.org/https://doi.org/10.1016/0034-4877(72)90011-0} {\bibfield
  {journal} {\bibinfo  {journal} {Reports on Mathematical Physics}\ }\textbf
  {\bibinfo {volume} {3}},\ \bibinfo {pages} {275} (\bibinfo {year}
  {1972})}\BibitemShut {NoStop}%
\end{thebibliography}%

\end{CJK*}
\end{document}